%% file: main.tex
\documentclass[11pt]{article}
\input{./Misc/Used_packages.tex}
\begin{document}

\title{Universal mean-field framework for SIS epidemics on networks, based on graph partitioning and the isoperimetric inequality}
\author{K. Devriendt\thanks{ Faculty of Electrical Engineering, Mathematics and
Computer Science, P.O Box 5031, 2600 GA Delft, The Netherlands; \emph{email}:
k.l.t.devriendt@student.tudelft.nl } ~and P. Van Mieghem}
\date{Delft University of Technology \\
\today
}
\maketitle

\begin{abstract}
We propose a new approximation framework that unifies and generalizes a number of existing mean-field approximation methods for the SIS epidemic model on complex networks. We derive the framework, which we call the Universal Mean-Field Framework (UMFF), as a set of approximations of the exact Markovian SIS equations. Our main novelty is that we describe the mean-field approximations from the perspective of the isoperimetric problem, which results in bounds on the UMFF approximation error. These new bounds provide insight in the accuracy of existing mean-field methods, such as the N-Intertwined Mean-Field Approximation (NIMFA) and Heterogeneous Mean-Field method (HMF) which are contained by UMFF. Additionally, the isoperimetric inequality relates the UMFF approximation accuracy to the regularity notions of Szemer\'edi's regularity lemma, which yields a prediction about the behavior of the SIS process on large graphs.
\end{abstract}
%-------------
\input{./Sections/Introduction_section.tex}
\input{./Sections/Background_the_SIS_epidemic_model.tex}
\input{./Sections/Definition_of_the_GHMF_framework.tex}
\input{./Sections/Derivation_of_the_GHMF_equations.tex}
\input{./Sections/Existing_mf_methods_contained_by_GHMF.tex}
\input{./Sections/The_Isoperimetric_Problem.tex}
\input{./Sections/EML_and_SRL.tex}
\input{./Sections/Related_work.tex}
\input{./Sections/Conclusion.tex}
\bibliographystyle{abbrv}
\bibliography{Bib.bib}
\section*{Appendices}
\appendix
\input{./Appendix/A_Kolmogorov_equations.tex}
\input{./Appendix/A_Higher-order_UMFF.tex}
\input{./Appendix/A_Proof_of_isoperimetric_inequalities.tex}

\end{document}

%% file: Misc/Used_packages.tex
\usepackage{amssymb}
\usepackage{amsfonts}
\usepackage{amsmath}
\usepackage{graphicx}%
\usepackage{mathtools} %(krl) for dcases
\providecommand{\U}[1]{\protect\rule{.1in}{.1in}}
\newtheorem{theorem}{Theorem}

\usepackage{tikz}
\usetikzlibrary{automata,positioning}
\usepackage{booktabs}
\usepackage{bbm}
\usetikzlibrary{cd}
\usepackage{enumerate}   
\newtheorem{problem}{Problem}
\newtheorem{definition}{Definition}

\newtheorem{approximation}{Approximation}
\newtheorem{lemma}{Lemma}

%% file: Sections/Introduction_section.tex
\section{Introduction}
The spread of epidemic diseases on complex networks is a widely studied topic in the field of network science \cite{Pastor-Satorras_2015}. While the mathematical study of epidemics dates back to the work of Bernoulli in the 18th century, the focus on the role of network topology only started at the end of the 20th century with the work of Kephart and White \cite{Kephart_1991}. With the recent observations that network structures seem ubiquitous in both natural and man-made systems, a better understanding of the interplay between dynamic processes and network structure has become an important pursuit. For the case of diseases, a better knowledge of the interaction between network features and the resulting spreading behavior could be crucial in managing epidemic outbreaks in the future. More generally, the theoretical study of spreading diseases is related to a much wider class of dynamic processes on networks like the spreading of information, computer viruses or opinions.
\\
In the study of epidemics on complex networks, the compartmental model of Kermack and McKendrick \cite{Kermack_1927} from 1927 is regarded as a basic disease model. In compartmental models, each entity in the population is assumed to be in a certain state, for instance healthy, contagious, immune or others. The state of each entity, from now on called node, can change based on the current state of the node itself and, in the case of nodes in a network, its neighboring nodes. By these local interactions the disease can spread, die out or show other behaviors depending on the model. A more general overview of the basic models and current progress in the field of epidemics on complex networks is given in \cite{Pastor-Satorras_2015}. 
\\
Like many network-epidemic studies, we will focus on one specific compartmental model: the SIS (Susceptible-Infected-Susceptible) model. The SIS model is often used because it is simple enough for a deep theoretical study while still being complex enough to exhibit global behavior that is non-trivially coupled to the small-scale process and the topology of the underlying network. In the SIS model, each node in the network can be in either of two states: susceptible (S) or infected (I). These states can change over time when an infected node is cured, or when a susceptible node is infected by a sick neighbor. These curing and infection events are stochastic processes that determine the dynamics of the disease. For a given initial distribution of infected nodes, the basic questions in studying the SIS model are then: what is the evolution of the state of the nodes in the network, how many nodes are infected in the metastable state, does the disease die out before reaching a significant fraction of the population, etc.
\\
To address these questions, some further assumptions need to be made about the dynamics of the SIS process. In the Markovian SIS model, on which this article focuses, the infection and healing events are modeled as Poisson processes. More general distributions are possible \cite{Cator_2013}, but with the Poisson assumption, the waiting times for infection or healing events are exponentially distributed which means that they satisfy the memoryless property, and the transitions between different configurations of the system become equivalent to state transitions in a Markov Chain. For specified rates of the Poisson processes, the evolution of the process can be exactly described based on Markov theory \cite{Van_Mieghem_2009}. However, since the number of possible states of the system grows exponentially with the number of nodes, this exact description is not practical. Consequently, several methods have been developed that approximate the SIS model in order to make analysis possible and investigate the interesting interaction between the process and the underlying topology. Notably, the N-Intertwined Mean-Field Approximation (NIMFA) \cite{Van_Mieghem_2009} and the Heterogeneous Mean-Field method (HMF) \cite{Pastor-Satorras_2001},\cite{Boguna_2002} are two widely-used approximation methods, which we will show to be contained by the introduced framework. An overview of these two methods and other SIS approximation methods can be found in \cite{Pastor-Satorras_2015} and \cite{Wang_2017}.
\\
In this article we present a general framework, called the Universal Mean-Field Framework (UMFF), to approximate the exact SIS description. The framework describes two general approximation steps that result in a set of approximate SIS equations (the UMFF equations). UMFF contains a number of existing mean-field methods, like NIMFA and HMF, and additionally extends the range of known SIS approximation methods. Apart from this unification and generalization, our main results are based on the close connection between the infection process in SIS epidemics and the well-studied isoperimetric problem \cite{blasjo_2005},\cite{Osserman_1978}. This connection provides novel insights, e.g. about the scaling behavior of the SIS process on large graphs, and allows us to deduce powerful bounds on the UMFF approximations.
\\
\\
In Section \ref{S_background}, we start by defining the SIS epidemic model on networks and elaborate on the feasibility of the exact SIS description. Then, in Section \ref{S_definition of GHMF}, we define UMFF, which consists of two approximations and the resulting UMFF equations. In Section \ref{S_derivation of GHMF}, we derive how the UMFF equations follow from the exact SIS equations by subsequently introducing the two approximations. 
In Section \ref{S_existing methods}, we describe how the existing mean-field methods are contained by UMFF. Section \ref{S_isoperimetric problem} introduces the isoperimetric problem and describes its analogy with the infection process. This analogy leads to the topological UMFF approximation and bounds on this approximation. In Section \ref{S_EML and SRL}, we discuss the relation between UMFF and Szemer\'edi's regularity lemma and explore the implications of this relation for the SIS process on large graphs. Section \ref{S_Related work} overviews some related work. Finally, Section \ref{S_conclusion} concludes the article by summarizing the main properties of UMFF and by suggesting some future research directions.

%% file: Sections/Background_the_SIS_epidemic_model.tex
\section{Background: the SIS epidemic model}
\label{S_background}
The SIS epidemic model is a compartmental model for modeling the spread of epidemic diseases on networks. We consider the process evolution on a specified network, which we represent by a graph $G(\mathcal{N},\mathcal{L})$, where $\mathcal{N}$ is the set of $N$ nodes and $\mathcal{L}$ the set of $L$ undirected, unweighted links between pairs of nodes. A convenient way of representing the graph structure is the $N\times N$ adjacency matrix $A$, with elements:
$$
a_{ij} = 
\begin{cases}
1 &\quad \text{if } (i,j)\in\mathcal{L} \\
0 &\quad \text{otherwise}
\end{cases}
$$
Since we consider undirected and unweighted graphs, the adjacency matrix $A$ is real and symmetric, possessing the following eigendecomposition:
$$
A = X\Lambda X^T = \sum_{i=1}^N\lambda_ix_ix_i^T
$$
where $X$ is the orthogonal eigen-matrix with eigenvectors $x_i$ as columns, and $\Lambda=\operatorname{diag}(\lambda_1,\lambda_2,\dots,\lambda_N)$ the diagonal matrix with eigenvalues on its diagonal. Because the adjacency matrix $A$ is real and symmetric, all eigenvalues are real and can be ordered as $\lambda_1\geq\lambda_2\geq\dots\lambda_N$. Another matrix capturing the graph structure is the Laplacian matrix $Q$, defined as:
$$
Q = \Delta - A
$$
where $\Delta$ is the diagonal matrix containing the node degrees. Since the Laplacian $Q$ is also real and symmetric matrix, we can write the eigendecomposition:
$$
Q = ZMZ^T = \sum_{i=1}^N\mu_iz_iz_i^T
$$
where $Z$ is the orthogonal eigen-matrix with eigenvectors $z_i$ as columns, and $M=\operatorname{diag}(\mu_1,\mu_2,\dots,\mu_N)$. Since all rows of $Q$ sum to zero, it holds that $Qu=0$, where $u$ is the all-one vector. The eigenvalue equation $Qu=\mu_Nu$ with $\mu_N=0$ illustrates that $Q$ has at least one zero eigenvalue, according to the eigenvector $\frac{u}{\sqrt{N}}$. The Laplacian $Q$ is positive semidefinite, which means that all eigenvalues are non-negative, i.e. $\mu_{i}\geq 0$ for all $i\leq N$. Additionally, the multiplicity of the zero eigenvalue $\mu_N$ is known to be one for connected graphs \cite{Van_Mieghem_GS}. Hence, for any connected graph, we can write the ordered sequence of Laplacian eigenvalues $0=\mu_N<\mu_{N-1}\leq\dots\leq\mu_1$.
\\
\\
The disease state of each node $n\in\mathcal{N}$ at a given time $t$, is captured by the variable:
$$
W_n(t) \in\left\lbrace 0,1\right\rbrace
$$ 
The expression $W_n(t)=0$ means that node $n$ is healthy, but susceptible (S) to the disease, while $W_n(t)=1$ means that the node is infected (I) and contagious. Since the SIS process is a stochastic process, $W_n(t)$ is a Bernoulli random variable and the infection probability of node $n$ equals $\Pr[W_n(t)=1]$. The evolution of the state probabilities over time is governed by the disease dynamics
$$S\rightarrow I\rightarrow S$$
which means that susceptible nodes can become infected nodes, which in turn can become susceptible. The $S\rightarrow I$ transition is called \textit{infection} and can occur when a susceptible node $n$ has an infected neighbor $j$ in the network. The $I\rightarrow S$ transition is called \textit{curing} and captures the process where each infected node has the possibility to be cured. To make the dynamics tractable, the infection and curing events are assumed to be independent Poisson processes. In particular, for the curing process,
\begin{equation} \label{eq_curing process}
\Pr[W_n(t+h)=0\vert W_n(t)=1] = \delta e^{-\delta h}
\end{equation}
means that, disregarding all other processes, the waiting time for the $I\rightarrow S$ transition is exponentially distributed with rate $\delta$. In general, each node $n$ can have a different, time-dependent rate $\delta_n(t)$, but further in this work we consider a fixed and time-independent rate $\delta$. If we consider just one link between a susceptible node $n$ and an infected node $j$, which we will call an infective link, then the infection process obeys
\begin{equation}\label{eq_infection process}
\Pr[W_n(t+h)=1\vert W_n(t)=0] = \beta e^{-\beta h}
\end{equation}
where the occurrence of other processes is ignored (which holds for $h\rightarrow 0$) and where we thus assume that the infected neighbor node $j$ stays infected and does not cure, i.e. $W_j(t+s)=1$ for $s\in[0,h]$. Again, each link $(n,j)\in\mathcal{L}$ can have a specific rate $\beta_{nj}(t)$, but for simplicity we assume a fixed and time-independent rate $\beta$. For notational purposes, we will often omit the time reference $t$ in time-dependent variables by writing $W_n$ instead of $W_n(t)$ and similarly for other time-dependent variables. It is also possible to model the infection and curing events as more general renewal processes \cite{Van_Mieghem_PA}, which results in different distributions for the waiting times \eqref{eq_curing process} and \eqref{eq_infection process}. For non-Poissonian processes, the Markov property no longer holds but approaches still exist to describe the SIS process \cite{Cator_2013}.
\\
\\
The expressions of the curing and infection processes show that the evolution of the process at a certain time, only depends on the state of the process at that time. This means that the system is memoryless and can effectively be described as a Markov process. The Kolmogorov equations and the infinitesimal generator of this continuous-time Markov Chain can then be deduced, based on the SIS dynamics \cite{Van_Mieghem_2009} (see also Appendix \ref{ASS_kolmogorov equations}). The infinitesimal generator describes the transition rates between the disease states of the system, which allows for an exact description of the Markov process for a given initial state. Unfortunately, there are $2^N$ possible states for an SIS process on an $N$ node network, which means that for roughly $N>20$, finding a solution of the $2^N$ linear equations becomes infeasible. This complexity of representing all possible disease states on a network is the main problem in describing the SIS process and especially, as we will show later, because of the dependence of the number of infective links (and thus the transition rates) on the full state information. An exact description of the process requires to calculate the probability $\Pr[W(t)=w]$ that the state vector $W(t)= [W_1(t),W_2(t),\dots,W_N(t)]^T$ equals a certain state vector $w$, for each possible state. Such a state $w$ is a zero-one Bernoulli vector, or $w\in\lbrace 0,1\rbrace^N$, which shows again that there are $2^N$ possible state vectors. 
\\
To resolve the complexity problem of the exact SIS equations, it is necessary to introduce approximations. The basic idea of approximating the SIS process lies in the description of the state by a different variable than the random variable $W(t)$, and to find the governing equations such that the exact dynamics are approximately described by that variable. UMFF relies on two different types of variables: the number of infected nodes $\widetilde{W}(t)$ (which is a random variable) and the (deterministic) expected number of infected nodes $\mathbf{E}[\widetilde{W}(t)]$. Apart from being a lower-dimensional description for the SIS process and thus addressing the exponential complexity problem, the number of infected nodes and the expected number of infected nodes are also more insightful variables.

%% file: Sections/Definition_of_the_GHMF_framework.tex
\section{Definition of the Universal Mean-Field Framework}
\label{S_definition of GHMF}
To describe UMFF, we need a number of preliminary definitions and notations. Firstly, we define a graph partitioning as follows:
\begin{definition}[Partitioning] A partitioning $\pi$ of graph $G$ defines a partitioning of the node-set $\mathcal{N}$ of $G$ into $K$ non-empty, disjoints partitions $\mathcal{N}_k\subseteq\mathcal{N}$ such that $\bigcup_{k=1}^{K}\mathcal{N}_k = \mathcal{N}$.
\end{definition}
By $N_k=\vert\mathcal{N}_k\vert$, we will denote the number of nodes in partition $k$, and by $L_{km}$, the number of links between nodes from partition $k$ and $m$ (and twice the number of links if $k=m$, see Table \ref{variable table}). \\
We will use the graph partitioning to group information of nodes belonging to the same partition, which results in a lower-dimensional description of the disease state and thus of the SIS process. A crucial concept of UMFF is to group nodes according to a partitioning $\pi$ and to consider the $K\times 1$ \textit{reduced-state vector} $\tilde{w}$ instead of the $N\times 1$ full-state vector $w$. The entry $\tilde{w}_k$ captures how many nodes are infected in partition $k$, which means that we have the relation:
\begin{equation}
\tilde{w}_k = \sum_{i\in\mathcal{N}_k}w_i
\end{equation}
for each $k\in\lbrace{1,2,\dots,K}\rbrace$, where $\tilde{w}_k$ is bounded as $0\leq\tilde{w}_k\leq N_k$. The reduced-state vector $\tilde{w}$ contains less information about the disease state; it is a coarser description of the disease state than the full-state $w$. In other words, one reduced state $\tilde{w}$ can correspond to a number of different full states $w$ (see also Appendix \ref{ASS_state probability}).
A number of additional notations follow from the state reduction, as defined in Table \ref{variable table}.
\input{./Figures/table.tex}

Based on the notion of a reduced state $\tilde{w}$, UMFF is defined as:
\begin{definition}[Universal Mean-Field Framework]\label{def_UMFF} Consider a graph $G(\mathcal{N},\mathcal{L})$, an SIS epidemic process with rates $(\beta,\delta)$ and a partitioning $\pi$ of the nodes into $K$ partitions. The UMFF equations are approximate equations for $\mathbf{E}[\widetilde{W}]$, the expected number of infected nodes in partition $k$:
\begin{equation} \label{eq_UMFF}
\frac{d\mathbf{E}[\widetilde{W}_k]}{dt} \approx -\delta\mathbf{E}[\widetilde{W}_k] + \beta\sum_{m=1}^K\tilde{a}_{km}\left(N_k-\mathbf{E}[\widetilde{W}_k]\right)\mathbf{E}[\widetilde{W}_m]
\end{equation}
\end{definition}
The UMFF equations follow from simplifying the exact SIS process description, using two approximations:
\begin{approximation}[Topological approximation] 
The number of infective links between susceptible nodes in partition $k$ and infected nodes in partition $m$ are approximated by:
\begin{equation} \label{approx_topological approximation}
(u-w)^TA^{(km)}w \approx (\tilde{u}-\tilde{w})^T\widetilde{A}^{(km)}\tilde{w} = \tilde{a}_{km}(N_k-\tilde{w}_k)\tilde{w}_m
\end{equation}
\end{approximation}
\textit{Remark:} The relations
$$
(u-w)^TA^{(km)}w = \sum_{i=1}^N\sum_{j=1}^Na^{(km)}_{ij}(1-w_i)w_j = \sum_{i\sim j}\mathbbm{1}_{\lbrace{(1-w_{i\in\mathcal{N}_k})w_{j\in\mathcal{N}_m}}\rbrace}
$$
where $\sum_{i\sim j}$ runs over all links $(i,j)\in\mathcal{L}$, show that $(u-w)^TA^{(km)}w$ indeed equals the number of infective links between susceptible nodes in partition $k$ and infected nodes in partition $m$.
\begin{approximation}[Moment-closure approximation] 
The covariance between the random variables $\widetilde{W}_k$ and $\widetilde{W}_m$ is approximated by zero:
\begin{equation} \label{approx_moment-closure approximation}
\operatorname{Cov}[\widetilde{W}_k,\widetilde{W}_m] \approx 0 \Rightarrow \mathbf{E}[\widetilde{W}_k\widetilde{W}_m]\approx\mathbf{E}[\widetilde{W}_k]\mathbf{E}[\widetilde{W}_m]
\end{equation}
\end{approximation}
In the next section, we show how the UMFF equations are found from the exact SIS  process description subject to approximations \eqref{approx_topological approximation} and \eqref{approx_moment-closure approximation}. The idea behind the topological approximation is further discussed in Section \ref{S_isoperimetric problem}, while the moment-closure approximation is addressed in Appendix \ref{A_higher-order UMFF}.

%% file: Figures/table.tex
\begin{table}[h!]
  \centering
    \label{variable table}
  \begin{tabular}{ccc}
    \toprule
    & Single node & Partition $(\pi)$\\
    \midrule
    \midrule
    \textbf{Node/Partition indicator}	& Node $i\in\lbrace 0,1,\dots,N\rbrace$ & Partition $k\in\lbrace 0,1,\dots,K\rbrace$ 
    \\ 
\addlinespace[0.5em]
    \textbf{Indicator vector} 			& $e_i\in\mathbb{R}^N$ & $\tilde{e}_k\in\mathbb{R}^K$
    \\
          									& $(e_i)_j = \mathbbm{1}_{\lbrace{i=j}\rbrace}$ & $(\tilde{e}_k)_m=\mathbbm{1}_{\lbrace{k=m}\rbrace}$
     \\
\addlinespace[0.5em]
     \textbf{Partition sum-vector} 		& $s_k\in\mathbb{R}^N$ & N.A. 
     \\
    									& $(s_k)_i = \mathbbm{1}_{\lbrace{i\in\mathcal{N}_k}\rbrace}$ &
	 \\
\addlinespace[0.5em]
     \textbf{All-one vector}			& $u=({1,1,\dots,1})^T$ & $\tilde{u} = (N_1,N_2,\dots,N_K)^T$ 
     \\
\addlinespace[0.5em]
     \textbf{State vector}				& $w=(w_1,w_2,\dots,w_N)^T$ & $\tilde{w}=(\tilde{w}_1,\tilde{w}_2,\dots,\tilde{w}_k)^T$
     \\
      										& $w_i=\mathbbm{1}_{\lbrace{\text{node i is infected}}\rbrace}$ & $\tilde{w}_k=s_k^Tw$
	 \\ 
\addlinespace[0.5em]
     \textbf{Adjacency matrix}			& $A\in\mathbb{R}^{N\times N}$ & $\widetilde{A}\in\mathbb{R}^{K\times K}$
     \\
    									& $a_{ij}=\mathbbm{1}_{\lbrace{(i,j)\in\mathcal{L}}\rbrace}$ & $\widetilde{a}_{km} = \frac{s_k^TAs_m}{N_kN_m} = \frac{L_{km}}{N_kN_m}$
	 \\
\addlinespace[0.5em]
     \textbf{Submatrix} 				& $A^{(km)}$ & $\widetilde{A}^{(km)}$ 
     \\
   										& $a^{(km)}_{ij} = a_{ij}\mathbbm{1}_{\lbrace{i\in\mathcal{N}_k\text{ and }j\in\mathcal{N}_m}\rbrace}$ & $\tilde{a}^{(km)}_{ij} = \tilde{a}_{ij}\mathbbm{1}_{\lbrace{i=k\text{ and }j=m}\rbrace}$
	 \\     
    \bottomrule
  \end{tabular}
  \caption{Overview of node-level and partition-level variables according to a specific partitioning. $\mathbbm{1}$ is the indicator function for which $\mathbbm{1}_{\lbrace{S}\rbrace}=1$ if statement $S$ is true and zero otherwise.}
\end{table}

%% file: Sections/Derivation_of_the_GHMF_equations.tex
\section{Derivation of the Universal Mean-Field Framework}
\label{S_derivation of GHMF}
Figure \ref{fig_variables} overviews the variables and approximations involved in UMFF, and how the UMFF equations are derived from the exact SIS equations. Additionally, 
it shows for which particular choices of partitioning, UMFF is to equivalent to existing mean-field methods (see also Section \ref{S_existing methods}). For $K=N$ partitions, each partition consists of exactly one node.
\\
\input{./Figures/variables.tikz}
In the next sections, we follow the variables in Figure \ref{fig_variables} from left to right to describe the derivation of the UMFF equations.
%----------------------------------------------------------
\subsection{Exact SIS equations}
\label{SS_derivation exact}
The UMFF approximation of the SIS process is based on two process variables: the reduced-state probability $\Pr[\widetilde{W}(t) = \tilde{w}]$ for each reduced state $\tilde{w}$, and the expected number of infected nodes $\mathbf{E}[\widetilde{W}_k(t)]$ for each partition $k$. In Appendix \ref{ASS_state probability}, the reduced-state probabilities are derived as:
\begin{equation} \label{eq_state prob exact}
\begin{split}
\frac{d\Pr[\widetilde{W}=\tilde{w}]}{dt} =
&-\delta\sum_{k=1}^K \tilde{w}_k\Pr[\widetilde{W}=\tilde{w}] + \delta\sum_{k=1}^K(\tilde{w}_k+1)\Pr[\widetilde{W}=\tilde{w}+\tilde{e}_k] \\
&-\beta\sum_{k=1}^K\sum_{m=1}^K \sum_{w\in\mathcal{W}^k_{\tilde{w}_k}\cap\mathcal{W}^m_{\tilde{w}_m}} (u-w)^TA^{(km)}w\Pr[W=w] \\
&+\beta\sum_{k=1}^K\sum_{m=1}^K \sum_{w\in\mathcal{W}^k_{(\tilde{w}_k-1)}\cap\mathcal{W}^m_{\tilde{w}_m}}(u-w)^TA^{(km)}w\Pr[W=w]
\end{split}
\end{equation}
for any reduced-state vector $\tilde{w}$, where $\mathcal{W}^k_{x}= \left\lbrace w\in\lbrace 0,1\rbrace^N\vert w^Ts_k=x \right\rbrace$ is the set of all full states $w$ with $x$ infected nodes in partition $k$. 
\\
In Appendix \ref{ASS_expected number}, the equations for the expected number of infected nodes is derived as:
\begin{equation} \label{eq_prevalence exact}
\frac{d\mathbf{E}[\widetilde{W}_k]}{dt} = -\delta\mathbf{E}[\widetilde{W}_k] + \beta\sum_{m=1}^K\sum_{\tilde{w}_k=0}^{N_k}\sum_{\tilde{w}_m=0}^{N_m}\sum_{w\in\mathcal{W}^k_{\tilde{w}_k}\cap\mathcal{W}^{m}_{\tilde{w}_m}}(u-w)^TA^{(km)}w \Pr[W=w]
\end{equation}
for each partition $k$.
%----------------------------------------------------------
\subsection{Death-birth process}
\label{SS_derivation death-birth}
In Appendix \ref{ASS_state probability} is shown that the rate of the infection transitions $\tilde{w}\rightarrow\tilde{w}+\tilde{e}_k$ and $\tilde{w}-\tilde{e}_k\rightarrow\tilde{w}$, which are transitions resulting from any node in partition $k$ being infected, depends on the number of infective links. The consequence is that equation \eqref{eq_state prob exact} for the reduced-state probability $\Pr[\widetilde{W}=\tilde{w}]$ depends on the full-state probability $\Pr[W=w]$, which means that equations \eqref{eq_state prob exact} are not a closed set of equations. 
\\
This closure problem is solved by invoking the UMFF topological approximation \eqref{approx_topological approximation}
$$
(u-w)^TA^{(km)}w \approx (\tilde{u}-\tilde{w})^T\widetilde{A}^{(km)}\tilde{w}
$$
which enables the simplifications 
\begin{equation}\label{simpl1}
\begin{dcases}
\sum_{w\in\mathcal{W}^k_{\tilde{w}_k}\cap\mathcal{W}^m_{\tilde{w}_m}}(u-w)^TA^{(km)}w\Pr[W=w] \approx (\tilde{u}-\tilde{w})^T\widetilde{A}^{(km)}\tilde{w}\Pr[\widetilde{W}_k=\tilde{w}_k,\widetilde{W}_m=\tilde{w}_m]
\\
\sum_{w\in\mathcal{W}^k_{(\tilde{w}_k-1)}\cap\mathcal{W}^m_{\tilde{w}_m}}(u-w)^TA^{(km)}w\Pr[W=w] \approx (\tilde{u}-(\tilde{w}-\tilde{e}_k))^T\widetilde{A}^{(km)}\tilde{w}\Pr[\widetilde{W}_k=\tilde{w}_k-1,\widetilde{W}_m=\tilde{w}_m]
\end{dcases}
\end{equation}
to be made in equation \eqref{eq_state prob exact}. Filling in \eqref{simpl1} in the exact equations \eqref{eq_state prob exact} yields:
\begin{equation} \label{eq_state prob topological approximation 1}
\begin{split} 
\frac{d\Pr[\widetilde{W} = \tilde{w}]}{dt} \approx
&-\delta\sum_{k=1}^K\tilde{w}_k\Pr[\widetilde{W}=\tilde{w}] + \delta\sum_{k=1}^K(\tilde{w}_k+1)\Pr[\widetilde{W}=\tilde{w}+\tilde{e}_k] \\
&-\beta\sum_{k=1}^K\sum_{m=1}^K(\tilde{u}-\tilde{w})\widetilde{A}^{(km)}\tilde{w}\Pr[\widetilde{W}_k=\tilde{w}_k,\widetilde{W}_m=\tilde{w}_m] \\
&+\beta\sum_{k=1}^K\sum_{m=1}^K(\tilde{u}-(\tilde{w}-\tilde{e}_k))\widetilde{A}^{(km)}\tilde{w}\Pr[\widetilde{W}_k=\tilde{w}_k-1,\widetilde{W}_m=\tilde{w}_m]
\end{split}
\end{equation}
which no longer depends on the full-state probability $\Pr[W=w]$. While equation \eqref{eq_state prob topological approximation 1} has a cumbersome form, it is a closed set of equations that completely characterizes $\Pr[\tilde{W}(t)=\tilde{w}]$ for a given initial distribution $\Pr[\tilde{W}(0)=\tilde{w}]$.
\\
Moreover, since only transitions of the form $\tilde{w}\rightarrow\tilde{w}\pm\tilde{e}_k$ and $\tilde{w}\pm\tilde{e}_k\rightarrow\tilde{w}$ exist (i.e. only single nodes are infected or cured during one event), equation \eqref{eq_state prob topological approximation 1} is equivalent to the description of a K-dimensional death-birth process. The reduced-state vector $\tilde{w}$, which can be seen as a coordinate in an $(N_1+1)\times(N_2+1)\times\dots\times(N_K+1)$ lattice, is then the counting variable in $K$ dimensions for this death-birth process. Furthermore, equation \eqref{eq_state prob topological approximation 1} indicates that the birth rates are quadratic in $\tilde{w}$ and the death rates are linear in $\tilde{w}$, which means that the SIS process is equivalent to a higher-dimensional quadratic death-birth process. While no analytical solutions exist for the quadratic death-birth process \cite{Van_Mieghem_2017}, the equivalence between the SIS and the quadratic death-birth process is an interesting observation. It means that insights in one setting translate directly to the other (see also Section \ref{S_conclusion}).
%----------------------------------------------------------
\subsection{UMFF equations}
\label{SS_derivation UMFF}
The exact equations \eqref{eq_prevalence exact} for the expected number of infected nodes $\mathbf{E}[\widetilde{W}_k]$ are not "closed" for two reasons: the exact SIS dynamics depend on the number of infective links (i.e. on full-state probability $\Pr[W=w]$) and on higher-order moments, i.e. the first-order moment equations \eqref{eq_prevalence exact} depend on the second-order moments $\mathbf{E}[\widetilde{W}_k\widetilde{W}_m]$ (see also Appendix \ref{A_higher-order UMFF}).
Similar to the derivation of the death-birth process, invoking the UMFF topological approximation \eqref{approx_topological approximation} results in simplifications \eqref{simpl1}, which allows equation \eqref{eq_prevalence exact} to be approximated by:
\begin{equation}\label{eq_prevalence_approx1}
\frac{d\mathbf{E}[\widetilde{W}_k]}{dt} = -\delta\mathbf{E}[\widetilde{W}_k] + \beta\sum_{m=1}^K\sum_{\tilde{w}_k=0}^{N_k}\sum_{\tilde{w}_m=0}^{N_m}(\tilde{u}-\tilde{w})^T\widetilde{A}^{(km)}\tilde{w}\Pr[\widetilde{W}_k=\tilde{w}_k,\widetilde{W}_m=\tilde{w}_m]
\end{equation}
While the dependence on the full-state probability $\Pr[W=w]$ is solved in equation \eqref{eq_prevalence_approx1}, it still contains higher-order moment terms:
\begin{equation} \label{eq_definition moment}
\sum_{\tilde{w}_k=0}^{N_k}\sum_{\tilde{w_m}=0}^{N_m}\tilde{w}_k\tilde{w}_m\Pr[\widetilde{W}_k=\tilde{w}_k,\widetilde{W}_m=\tilde{w}_m] = \mathbf{E}[\widetilde{W}_k\widetilde{W}_m]
\end{equation}
for partition pairs $(k,m)$. In general, these second-order moments $\mathbf{E}[\widetilde{W}_k\widetilde{W}_m]$ cannot be determined from $\mathbf{E}[\widetilde{W}_k]$ and $\mathbf{E}[\widetilde{W}_m]$ alone. Invoking the UMFF moment-closure approximation \eqref{approx_moment-closure approximation}
$$
\operatorname{Cov}[\widetilde{W}_k,\widetilde{W}_m]\approx 0\Rightarrow \mathbf{E}[\widetilde{W}_k\widetilde{W}_m]\approx\mathbf{E}[\widetilde{W}_k]\mathbf{E}[\widetilde{W}_m]
$$
solves this closure problem by enabling equation \eqref{eq_prevalence_approx1} to be approximated by 
$$
\frac{d\mathbf{E}[\widetilde{W}_k]}{dt} \approx -\delta\mathbf{E}[\widetilde{W}_k]+\beta\sum_{m=1}^K\tilde{a}_{km}(N_k-\mathbf{E}[\widetilde{W}_k])\mathbf{E}[\widetilde{W}_m]
$$
which are the UMFF equations \eqref{eq_UMFF}. 
\\
In Appendix \ref{A_higher-order UMFF}, an extension of the UMFF equations for higher-order moments is described. These higher-order equations are more general, but a detailed description is not the focus of this article. 
%----------------------------------------------------------
\subsubsection*{Bounds on the moment-closure approximation}
For the particular case of $K=N$ partitions (for which UMFF is equivalent to NIMFA, see Section \ref{S_existing methods}), the infection probabilities of nodes are non-negatively correlated \cite{Cator_2014}, i.e. $\operatorname{Cov}[\widetilde{W}_k,\widetilde{W}_m]\geq 0$. Based on the definition of the covariance
\begin{equation}\label{eq_definition covariance}
\operatorname{Cov}[\widetilde{W}_k,\widetilde{W}_m] = \mathbf{E}[\widetilde{W}_k\widetilde{W}_m]-\mathbf{E}[\widetilde{W}_k]\mathbf{E}[\widetilde{W}_m]
\end{equation}
we can rewrite the exact equation \eqref{eq_prevalence_approx1} as:
\begin{equation} \label{eq_prevalence_cov}
\frac{d\mathbf{E}[\widetilde{W}_k]}{dt} = -\delta\mathbf{E}[\widetilde{W}_k]+\beta\sum_{m=1}^K\tilde{a}_{km}(N_k-\mathbf{E}[\widetilde{W}_k])\mathbf{E}[\widetilde{W}_m] - \beta\sum_{m=1}^N\tilde{a}_{km}\operatorname{Cov}[\widetilde{W}_k,\widetilde{W}_m]
\end{equation}
Omitting the negative term $-\tilde{a}_{km}\operatorname{Cov}[\widetilde{W}_k,\widetilde{W}_m]$ in equation \eqref{eq_prevalence_cov} implies that for $K=N$ partitions, the moment-closure approximation is an upper-bound of the true process. However, for any other partitioning ($K\neq N$) we do not know about any such results for $\operatorname{Cov}[\widetilde{W}_k,\widetilde{W}_m]$. In other words, we do not know how to bound the UMFF moment-closure approximation error. 

%% file: Figures/variables.tikz
%Source: http://steventhornton.ca/markov-chains-in-latex/
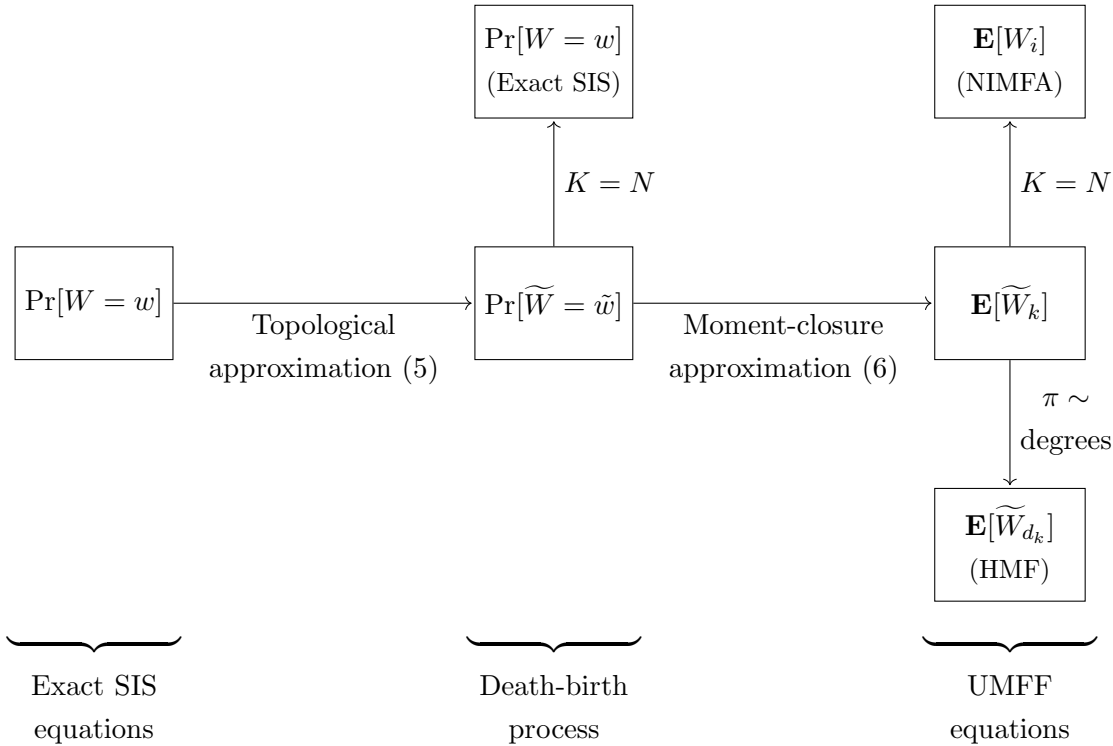
\begin{figure}[h!]
\label{fig_variables}
\centering
\begin{tikzpicture}[node distance = 4cm]
\tikzstyle{block} = [draw, rectangle,minimum width = 2cm, minimum
height = 1.5cm]

\tikzstyle{vecArrow} = [thick, decoration={markings,mark=at position
   1 with {\arrow[semithick]{open triangle 60}}},
   double distance=1.4pt, shorten >= 5.5pt,
      postaction = {draw,line width=1.4pt, white,shorten >= 4.5pt}]
\tikzstyle{innerWhite} = [semithick, white,line width=1.4pt, shorten >= 4.5pt]

        % Draw the states
        \node[block] (w1) {$\Pr[W=w]$};
        \node[block,right=of w1] (w2) {$\Pr[\widetilde{W}=\tilde{w}]$};
        \node[block,right=of w2] (w5) {$\mathbf{E}[\widetilde{W}_k]$};        
        
        \node[block, above=1.7cm of w2, align=center] (w3) {$\Pr[W=w]$\\\small(Exact SIS)};
        \node[block, above=1.7cm of w5, align=center] (w6) {$\mathbf{E}[W_i]$\\\small(NIMFA)};
        
        \node[block, below=1.7cm of w5, align=center] (w7) {$\mathbf{E}[\widetilde{W}_{d_k}]$\\\small(HMF)};
                
                \node[draw=none, below=3.5cm of w1,align=center] (w12) {$\underbrace{\hspace{6em}}_{}$\\Exact SIS\\equations};
        \node[draw=none, below=3.5cm of w2,align=center] (w11) {$\underbrace{\hspace{6em}}_{}$\\Death-birth\\process};
        \node[draw=none, below=3.5cm of w5,align=center] (w13) {$\underbrace{\hspace{6em}}_{}$\\UMFF\\equations};        
 		
        % Connect the states with arrows
        \draw[every loop]
        (w1) edge[auto=right, align=center] node {Topological \\approximation \eqref{approx_topological approximation}} (w2)        
        (w2) edge[auto=right, align=center] node {Moment-closure \\approximation \eqref{approx_moment-closure approximation}} (w5)
        (w2) edge[auto=right] node {$K=N$} (w3)
        %(w2) edge[auto=left] node {} (w4)
        (w5) edge[auto=right] node {$K=N$} (w6)
        (w5) edge[auto=left,align = center] node {$\pi\sim$\\degrees} (w7)
        
        ;
        %(w1) edge[bend right, auto=left] node {$r_{12}$} (w2)
            
\end{tikzpicture}
\caption{Schematic representation of the relationship between the different variables involved in the UMFF approximation steps.}
\end{figure}

%% file: Sections/Existing_mf_methods_contained_by_GHMF.tex
\section{Existing mean-field methods contained by UMFF}
\label{S_existing methods}
An important feature of UMFF is that by particular choices of graph partitioning, the UMFF equations are equivalent to existing mean-field methods. In particular, the widely-used N-Intertwined Mean-Field Approximation \cite{Van_Mieghem_2009} and Heterogeneous Mean-Field approximation \cite{Pastor-Satorras_2001} are contained by UMFF. Additionally, by the higher-order extension of UMFF described in Appendix \ref{A_higher-order UMFF}, also second-order NIMFA \cite{Cator_2014} and pair Quenched Mean-Field theory \cite{Mata_2013} are contained by (higher-order) UMFF.
%--------------------------------------------------
\subsection{N-Intertwined Mean-Field Approximation (NIMFA)}
The N-Intertwined Mean-Field Approximation \cite{Van_Mieghem_2009} incorporates the full topological information of the graph in its system of equations. The only approximation consists of assuming independence between the infection states of adjacent nodes. Denoting the infection probability of node $k$ by $\rho_k=\Pr[W_k=1]$, the NIMFA equations for $1\leq k\leq N$ are given by \cite{Van_Mieghem_2009}:
\begin{equation} \label{NIMFA ode l}
\frac{d\rho_k}{dt} = -\delta \rho_k + \sum_{m=1}^{N}\beta a_{km} (1-\rho_k)\rho_m,
\end{equation}
The same NIMFA equations \eqref{NIMFA ode l} are retrieved from UMFF with $K=N$ partitions, which corresponds to each node being in a separate partition. The expected number of infected nodes in a partition $\mathbf{E}[\widetilde{W}_k]$ is then equal to the infection probability $\rho_k$ of node $k$ that makes up that partition. For this partitioning, we have $N_k=1$ and $\widetilde{A}=A$, illustrating that the NIMFA equations \eqref{NIMFA ode l} are indeed a particular case of the UMFF equations \eqref{eq_UMFF}.
%--------------------------------------------------
\subsection{Heterogeneous mean-field method (HMF)}
Pastor-Satorras and Vespignani \cite{Pastor-Satorras_2001} introduced the Heterogeneous Mean-Field method, which approximates the SIS process based on the assumption that all nodes of a certain degree are equivalent (in their connections with other nodes). Consequently, the SIS process is described based on the degree distribution of the underlying graph.
\\
Different from UMFF and NIMFA, HMF \cite{Pastor-Satorras_2001} does not assume a known graph $G$, but rather considers a class of graphs. Specifically, in HMF the epidemic is assumed to take place on a graph with a specified degree distribution and with the link probability between pairs of nodes independent of their degrees. For each degree $d_1\leq d_k\leq d_K$, the probability distribution $\Pr[D=d_k]$ denotes the probability that a randomly chosen node has degree $d_k$. The variable $0\leq \tilde{\rho}_k\leq1$ reflects the expected fraction of infected nodes with degree $d_k$:
\begin{equation} \label{HMF}
\frac{d\tilde{\rho}_k}{dt}=-\delta \tilde{\rho}_k + \beta k(1-\tilde{\rho}_k)\Theta
\end{equation}
where $\Theta$ is the probability that a healthy node is linked to an infected node. The value of $\Theta$ is calculated in \cite{Pastor-Satorras_2001}, based on the connection probability of nodes of degree $d_k$ to infected nodes in the rest of the network, as:
\begin{equation} \label{eq_theta}
\Theta=\sum_{m=1}^{K}\tilde{\rho}_m\frac{d_m\Pr[D=d_m]}{\sum_{i=1}^{K}d_i\Pr[D=d_i]}
\end{equation}
Substituting expression \eqref{eq_theta} for $\Theta$ in \eqref{HMF} gives:
\begin{equation} \label{HMF explicit}
\frac{d\tilde{\rho}_k}{dt}=-\delta \tilde{\rho}_k + \beta\sum_{m=1}^{K}\frac{d_kd_m\Pr[D=d_m]}{\sum_{i=1}^Kd_i\Pr[D=d_i]}(1-\tilde{\rho}_k)\tilde{\rho}_m
\end{equation}
Introducing the variable $\rho_k=\Pr[D=d_k]\tilde{\rho}_k$ then yields:
\begin{equation} \label{HMF explicit scaled}
\frac{d\rho_k}{dt}=-\delta \rho_k + \beta\sum_{m=1}^{K}\frac{d_kd_m}{\sum_{i=1}^Kd_i\Pr[D=d_i]}\left(\Pr[D=d_k]-\rho_k\right)\rho_m
\end{equation}
While the above equations \eqref{HMF explicit scaled} are derived in HMF for a probabilistic graph, the same equations are found from UMFF for a particular graph with the same degree distribution, namely $N_k=c\Pr[D=d_k]$ nodes of degree $d_k$ for some scalar $c\in\mathbb{R}$, and degree-uncorrelated links. For such a  graph, the number of links $L_{km} = s_k^TAs_m$ between nodes of degree $d_k$ and degree $d_m$ obeys the consistency relation $\sum_{m=1}^KL_{km} = N_kd_k$ as:
$$L_{km} = \frac{d_kd_mN_kN_m}{\sum_{i=1}^Kd_iN_i},$$ from which the UMFF equations follow as:
\begin{equation} \label{GHMF as HMF}
\frac{d\mathbf{E}[\widetilde{W}_k]}{dt} = -\delta\mathbf{E}[\widetilde{W}_k] +\beta\sum_{m=1}^K\frac{d_kd_m}{\sum_{i=1}^Kd_iN_i}\left(N_k-\mathbf{E}[\widetilde{W}_k]\right)\mathbf{E}[\widetilde{W}_m]
\end{equation}
Equations \eqref{GHMF as HMF} are equivalent to \eqref{HMF explicit scaled} for the scaling $\mathbf{E}[\widetilde{W}_k]=c\rho$, where $c$ is the same scalar relating $N_k$ to $\Pr[D=d_k]$. Hence, the HMF equations are found from the UMFF framework by considering a specific graph realization consistent with the random graph properties assumed by HMF.
\\
Since HMF is a particular case of UMFF, HMF implicitly uses the UMFF moment-closure approximation \eqref{approx_moment-closure approximation} with respect to the partitioning according to node degree. As discussed in Section \ref{SS_derivation UMFF}, this means that we do not know whether the HMF equations give an upper-bound or a lower-bound on the infection probabilities, or how they relate to the exact SIS process in general.
\\
\\
Bogu\~n\'a and Pastor-Satorras \cite{Boguna_2002} extend the HMF model to random graphs with correlated degrees. Instead of only assuming $\Pr[D=d_k]$, also the probability $\Pr[i\sim j\vert i\in\mathcal{N}_k,j\in\mathcal{N}_m]$ that a node $i$ of degree $d_k$ links with a node $j$ of degree $d_m$ is assumed to be known for any pair of degrees $(d_k,d_m)$. With these extra assumptions in the HMF methodology, the SIS process is then approximately described based on the degree distribution and the linking probabilities. If we now consider a specific graph realization with $N_k=c_1\Pr[D=d_K]$ nodes of degree $d_k$ and with $L_{km}=c_2\Pr[i\sim j\vert i\in\mathcal{N}_k,j\in\mathcal{N}_m]$ links between nodes with degree $d_k$ and $d_m$ (for some scalars $c_1,c_2\in\mathbb{R}$), then again the UMFF equations \eqref{eq_UMFF} are equivalent to the correlated HMF equations. 
\\
In the same way that the HMF equations are fully determined by the degree distribution and the linking probabilities, also the UMFF equations are fully determined by $N_k$ and $L_{km}$. A consequence of the equivalence between UMFF \eqref{eq_UMFF} and (correlated) HMF \eqref{HMF explicit scaled}, is that we can bound the topological approximation errors of HMF (with respect to a specific realization of the probabilistic graph model).
\\
\\
Since the partitions $\mathcal{N}_k$ do not need to correspond to node degrees specifically, UMFF enables the SIS dynamics to be described for a wider range of graph classes. For any graph model, where a probability distribution $\Pr[K=k]$ of a graph belonging to partition $\mathcal{N}_k$ is given, together with a linking probability $\Pr[i\sim j\vert i\in\mathcal{N}_k,j\in\mathcal{N}_m]$, the UMFF equations can be directly found. Such graph models are more general than graphs with degree-based partitions only and, in some settings, specific structure in the graph might suggest a natural way to partition the nodes such that grouped nodes have a similar connectivity to the rest of the network (see also further directions in Section \ref{S_conclusion}).
%-----------------------------------------------------
\subsection{Second-order NIMFA and Pair Quenched Mean-Field theory}
\label{SS_sNIMFA and pQMF}
Second-order NIMFA (sNIMFA) \cite{Cator_2012} and Pair Quenched Mean-Field theory (pQMF) \cite{Mata_2013} are second-order mean-field methods developed to approximate the description of SIS dynamics on networks. 
\\
As described in Appendix \ref{A_higher-order UMFF}, sNIMFA is a second-order extension of NIMFA, approximating the joint probability $\Pr[W=w]$ by first and second-order moments $\mathbf{E}[W_i]$ and $\mathbf{E}[W_iW_j]$ for all nodes $i\neq j$. Similarly, pQMF \cite{Mata_2013} is an extension of Quenched Mean-Field theory (QMF)\cite{Castellano_2010}, which is an SIS approximation method introduced to investigate the epidemic threshold. The extension QMF$\rightarrow$pQMF is conceptually the same as NIMFA$\rightarrow$sNIMFA, but a different moment-closure approximation approximates the third-order moments.
\\
Both sNIMFA as well as pQMF are contained by the higher-order UMFF equations \eqref{eq_higher-order UMFF def}, for $K=N$ partitions and order $n=2$ if the generic moment-closure approximation is chosen as in \cite{Cator_2014} and \cite{Mata_2013} respectively.

%% file: Sections/The_Isoperimetric_Problem.tex
\section{The \textit{Isoperimetric Problem} in SIS epidemics}
\label{S_isoperimetric problem}
In this section, we focus on the UMFF topological approximation:
$$
(u-w)^TA^{(km)}w \approx (\tilde{u}-\tilde{w})^T\widetilde{A}^{(km)}\tilde{w}
$$
We first describe how the closure problem of equations \eqref{eq_state prob exact} and \eqref{eq_prevalence exact} can be related to the isoperimetric problem. Then, we show how this analogy leads to approximation \eqref{approx_topological approximation} and bounds on the approximation error. 
%-------------------------------------
\subsection{The isoperimetric problem}
The isoperimetric problem is an ancient problem that has interested many mathematicians throughout history. For the most basic form of the isoperimetric problem, we cite Bl\r{a}sj\"{o} \cite{blasjo_2005}, who provides a broad historical and conceptual overview of the isoperimetric problem:
\begin{problem}[The isoperimetric problem] Among all figures in the plane with a given perimeter $L$, which one encloses the greatest area $A$?
\end{problem}
\begin{theorem}[The isoperimetric theorem] The solution to the isoperimetric problem is the circle of perimeter $L$.
\end{theorem}
\begin{theorem}[The isoperimetric inequality] \label{th_isoperimetric inequality} For all figures with a given perimeter $L$ and area $A$, it holds that $L^2-4\pi A\geq 0$ and equality only occurs for the circle.
\end{theorem}
While the question in problem 1 might seem simple, and its solution intuitive, it took until the $20$'th century to rigorously prove the isoperimetric theorem. After the extensive historical study of the isoperimetric problem in the $2$D plane, similar problems were studied in different geometric contexts. The basic interest in these problems always consisted of describing the relationship between the \textit{volume} and \textit{surface} of a certain object, leading to isoperimetric inequalities of the form:
\begin{equation} \label{eq_conceptual II}
\theta_{\text{min}} \leq f(\text{volume})+g(\text{surface}) \leq \theta_{\text{max}}
\end{equation}
For instance, Osserman \cite{Osserman_1978} describes isoperimetric inequalities in higher dimensions, on curved surfaces and on general Riemannian manifolds. The geometric context of interest for UMFF, is the study of the isoperimetric problem on graphs (see for instance \cite{Chung_1996}).
%-----------------------------------------------------
\subsection{Infective links and infected nodes: an isoperimetric analogy}
\begin{figure}
\centering
\includegraphics[scale=0.5]{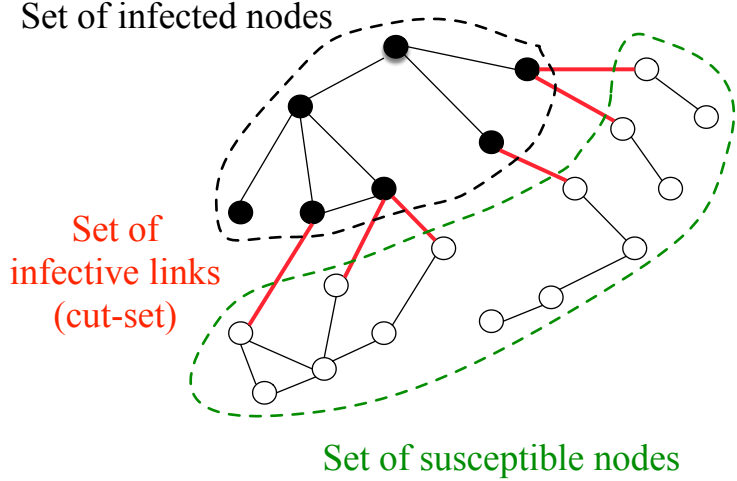}
\caption{Example of a disease state in a toy network. The infected and healthy nodes determine two separate partitions with the cut-set between them determining the set of infective links.}
\label{fig_cut-set}
\end{figure}

The dynamics of SIS epidemics are governed by two processes: infected are cured and infection takes place on infective links, i.e. the links between healthy and infected nodes.\emph{ The curing process is proportional to the number of infected nodes while the infection process is proportional to the number of infective links}. In a non-technical way we can associate the number of infected nodes to a volume on the graph, while the infective links accord to a surface or interface around the infected volume.\emph{ The curing process is then proportional to infected volume while the infection process is proportional to the infective surface}. This relation is illustrated in Figure \ref{fig_cut-set}, which represents a specific disease state on a toy-network. 
\\
To use the concepts of volume and surface in a more technical way, we must define a unit of volume and surface in the context of graphs: we define a set of one node to have unit volume, and a set of one link to have unit surface. Other choices are possible, e.g. the volume of a node being proportional to its degree, but for the purpose of deriving and bounding the UMFF topological approximation \eqref{approx_topological approximation}, this would be a less natural choice.
\\
\\
In the derivation of the exact reduced-state Kolmogorov equations \eqref{eq_state prob exact}, the transition rate between reduced states depends on the number of infective links. Hence, the exact equations \eqref{eq_state prob exact} for $\Pr[\widetilde{W}=\tilde{w}]$ and \eqref{eq_prevalence exact} for $\mathbf{E}[\widetilde{W}]$ are not closed, because they contain terms of the form $(u-w)^TA^{(km)}w\Pr[W=w]$. In the language of the isoperimetric problem, this \textit{closure problem} translates to the volume equations \eqref{eq_state prob exact} and \eqref{eq_prevalence exact} containing terms related to the surface. The UMFF topological approximation \eqref{approx_topological approximation} replaces the surface term by a function of volume terms and thus solves the closure problem. Now, by analogy with the isoperimetric problem, we can bound the approximation error caused by this replacement, as shown in Figure \ref{fig_conceptual diagram}, where $\epsilon$ represents the introduced error.
\\
\input{./Figures/Isoperimetric_analogy.tikz}
\\
It remains to find the correct translation of the isoperimetric inequality into the setting of SIS epidemics. The UMFF topological approximation is defined as \eqref{approx_topological approximation}:
$$
(u-w)^TA^{(km)}w \approx (\tilde{u}-\tilde{w})^T\widetilde{A}^{(km)}\tilde{w}
$$
which we can rewrite by introducing an error term $\epsilon\in\mathbb{R}$ as:
\begin{equation}
(u-w)^TA^{(km)}w = (\tilde{u}-\tilde{w})^T\widetilde{A}^{(km)}\tilde{w}+\epsilon
\end{equation}
or, by upper-bounding the error term $\vert\epsilon\vert\leq\theta$, as: 
\begin{equation}
\left\vert (u-w)^TA^{(km)}w - (\tilde{u}-\tilde{w})^T\widetilde{A}^{(km)}\tilde{w} \right\vert \leq \theta
\end{equation}
In the next subsection, we specify the error bound $\theta$ based on the isoperimetric inequalities on graphs. More than just providing an error bound, the analogy with the isoperimetric problem and the mathematical techniques used in the proofs (see Appendix \ref{A_proof isoperimetric}) also provide a motivation for the specific form of the UMFF topological approximation \eqref{approx_topological approximation}.
%-----------------------------------------------------
\subsection{Isoperimetric inequalities for the number of infective links}
The bound for the approximation error is based on the isoperimetric and discrepancy inequalities of Chung \cite{Chung_1996}:
\begin{theorem}[General-graph isoperimetric inequality] \label{topological approx theorem}
For a graph $G(\mathcal{N},\mathcal{L})$ and a partitioning $\pi$, the error of the UMFF topological approximation \eqref{approx_topological approximation} between any two partitions $k$ and $m$ is bounded as:
\begin{equation} \label{topological approx general graph}
\left\vert (u-w)^TA^{(km)}w - (\tilde{u}-\tilde{w})^T\widetilde{A}^{(km)}\tilde{w} \right\vert \leq \frac{\theta}{N}\sqrt{\tilde{w}_m(N-\tilde{w})(N_k-\tilde{w}_k)(N-(N_k-\tilde{w}_k))}
\end{equation}
where $\left\vert{\tilde{a}_{km}-\mu_i}\right\vert\leq\theta$ holds for $1\leq i< N$, with $\mu_i$ the eigenvalues of the Laplacian matrix based on $A^{(km)}$. \end{theorem}
For bi-regular graphs $A^{(km)}$, meaning that $A^{(km)}s_m = c_1s_m$ and $s_k^TA^{(km)}=c_2s_k^T$ for some constants $c_1,c_2\in\mathbb{R}$, a tighter bound can be given based on interlacing techniques of Haemers \cite{Haemers_1995}:
\begin{theorem}[Bi-regular-graph isoperimetric inequality] \label{bi-regular approx theorem}
For a graph $G(\mathcal{N},\mathcal{L})$ and a partitioning $\pi$ such that $A^{(km)}$ is bi-regular for some partitions $k$ and $m$, the error of the UMFF topological approximation \eqref{approx_topological approximation} is bounded as:
\begin{equation} \label{topological approx bi-regular graph}
\left\vert (u-w)^TA^{(km)}w - (\tilde{u}-\tilde{w})^T\widetilde{A}^{(km)}\tilde{w} \right\vert \leq \frac{\lambda_2}{N}\sqrt{\tilde{w}_k(N_k-\tilde{w}_k)\tilde{w}_m(N_m-\tilde{w}_m)}
\end{equation}
where $\lambda_2$ is the second-largest eigenvalue of $A^{(km)}$.
\end{theorem}
The proofs of Theorem \ref{topological approx theorem} and Theorem \ref{bi-regular approx theorem} are given in appendix \ref{A_proof isoperimetric} and rely heavily on proofs given by Chung \cite{Chung_1996} and Haemers \cite{Haemers_1995}.

%% file: Figures/Isoperimetric_analogy.tikz
%Source: http://steventhornton.ca/markov-chains-in-latex/
\begin{figure}[h!]
\centering
\begin{tikzpicture}[node distance=5cm]
\usetikzlibrary{arrows}
\tikzstyle{block} = [draw, rectangle,minimum width = 6cm, minimum
height = 0.9cm]
        % Draw the states
        \node[block, align = center] (1) {$\frac{d}{dt}(\text{\#I Nodes})\propto\text{\#I Links}$};
        \node[block, right=of 1, align = center] (2) {$\frac{d}{dt}(\text{\#I Nodes})\propto\text{\textbf{\#I Nodes}}+\epsilon$};
        \node[block, below=1cm of 1] (3) {$\frac{d}{dt}(\text{Volume})\propto\text{Surface}$};
        \node[block, below=1cm of 2] (4) {$\frac{d}{dt}(\text{Volume})\propto\text{\textbf{Volume}}+\epsilon$};
        \node[draw=none, above=0.5cm of 1] (5) {\underline{Exact SIS equations}};
        \node[draw=none, above=0.5cm of 2] (6) {\underline{UMFF equations}};
 
        % Connect the states with arrows
        \draw[every loop]
        %(1) edge[auto=left] node {\small Isoperimetric inequality \eqref{eq_conceptual II}} (2)
        (3) edge[auto=left] node {\small Isoperimetric inequality \eqref{eq_conceptual II}} (4)
        (1) edge[auto=left] node {\small Topological approximation \eqref{approx_topological approximation} } (2);
        \draw[latex'-latex',double]
        (1) edge[auto=left] node {\small analogy} (3)
        (2) edge[auto=left] node {\small analogy} (4);
        \end{tikzpicture}
\caption{Conceptual diagram depicting the analogy between the UMFF topological approximation \eqref{approx_topological approximation} and the isoperimetric inequality \eqref{eq_conceptual II}}. 
\label{fig_conceptual diagram}
\end{figure}
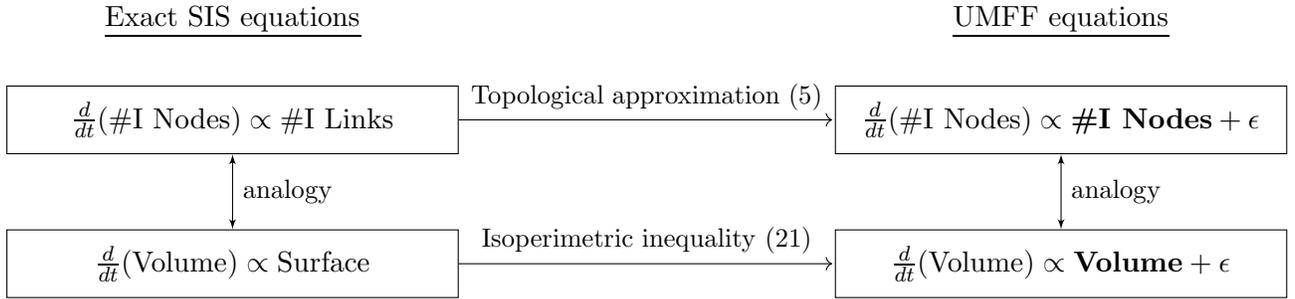

%% file: Sections/EML_and_SRL.tex
\section{UMFF and Szemer\'edi's regularity lemma}
\label{S_EML and SRL}
The isoperimetric problem is a well-studied mathematical problem that appears in many different fields, including graph theory and network science, and thus provides a conceptual link between those fields. For instance Szemer\'edi's regularity lemma (SRL) is a lemma with interesting implications for UMFF, which follows from the relation of both UMFF and SRL with the isoperimetric problem. 
We will discuss how SRL may indicate for which graphs the UMFF topological approximation \eqref{approx_topological approximation} is expected to be accurate, and for which the SIS dynamics are thus well approximated by the UMFF equations.
%-----------------------------------------------------------
\subsubsection*{Szemer\'edi's regularity lemma}
The following definitions and interpretations are based on Diestel's \cite{Diestel_2012} description of SRL. We start by defining a so-called regularity condition between pairs of partitions, which is related to the isoperimetric inequality.
\begin{definition}[$\epsilon$-regular partition pair]\label{e-regular partition pair}\cite{Diestel_2012}
Consider a graph $G(\mathcal{N},\mathcal{L})$ and two disjoint node partitions $\mathcal{N}_k,\mathcal{N}_m\subseteq\mathcal{N}$. If for any pair of subsets $\mathcal{N}_x\subseteq\mathcal{N}_k$ and $\mathcal{N}_y\subseteq\mathcal{N}_m$ of size $N_x$ and $N_y$ with $N_x\geq \epsilon N_k$ and $N_y\geq\epsilon N_m$ for some real $\epsilon>0$, the inequality
\begin{equation}\label{eq_e-regular}
\left\vert \frac{(u-s_x)^TA^{(km)}s_y}{N_xN_y} - \frac{s_k^TA^{(km)}s_m}{N_kN_m} \right\vert \leq \epsilon
\end{equation}
holds, then we say that the partition pair $(k,m)$ is $\epsilon$-regular.
\end{definition}
Inequality \eqref{eq_e-regular} can be rewritten as
\begin{equation} \label{eq_e-regular rewritten}
\left\vert (u-s_x)^TA^{(km)}s_y - \frac{L_{km}}{N_kN_m}N_xN_y\right\vert \leq \epsilon N_xN_y
\end{equation}
which shows that the regularity condition \eqref{eq_e-regular} is related to the difference between the size of the cut-set $(u-s_x)^TA^{(km)}s_y$ (for all subsets of partitions $k,m$ with $N_x,N_y$ nodes, respectively) and the approximate size of the cut-set: $\frac{L_{km}}{N_kN_m}N_xN_y$. For lower values of $\epsilon$, the regularity condition becomes stronger. Firstly, because the true size of the cut-set can deviate less from the approximate cut-set size if $\epsilon$ is smaller, and secondly because the regularity condition must hold for a larger range of subsets $(\mathcal{N}_x,\mathcal{N}_y)$, since $N_x\geq\epsilon N_k$ is a less stringent condition if $\epsilon$ is lower (and similarly for $\mathcal{N}_y$). 
\\
Based on the notion of $\epsilon$-regular partition pairs, we define a regularity condition on a partitioning $\pi$ of a graph:
\begin{definition}[$\epsilon$-regular graph partitioning]\cite{Diestel_2012}\label{e-regular graph partitioning} Consider a graph $G(\mathcal{N},\mathcal{L})$ with a partitioning $\pi$ of the nodes into $K+1$ partitions $\left\lbrace{\mathcal{N}_0,\mathcal{N}_1,\dots,\mathcal{N}_K}\right\rbrace$. Such a graph partitioning is called $\epsilon$-regular if it meets the following conditions:
\begin{enumerate}[(i)]
\item $N_0 \leq \epsilon N$
\item $N_1=N_2=\dots=N_K$
\item All except at most $\epsilon K^2$ of the partition pairs (k,m) for $1\leq k<m\leq K$ are $\epsilon$-regular
\end{enumerate}
\end{definition}
Roughly speaking, a graph partitioning is $\epsilon$-regular if it contains $K$ equally sized partitions (ii) such that most partition pairs are regular (iii), where one additional ``small" partition is allowed to exist (i) on which conditions (ii) and (iii) do not apply. For a given $K$, a smaller $\epsilon$ strengthens the regularity conditions. Firstly, because the regularity condition between partition pairs becomes stronger, secondly, because $N_0\leq \epsilon N$ means that a lower number of nodes are allowed to make up the ``leftover partition" $\mathcal{N}_0$ and, finally, because $\epsilon K^2$ becomes smaller, implying that an increasing proportion of the partition pairs need to satisfy the regularity condition \eqref{eq_e-regular}. Since condition (iii) holds for partition pairs $(k,m)$ with $k\neq m$, the regularity conditions only applies to links between partitions and not within partitions. \\
Based on the regularity notion of a graph partitioning, Szemer\'edi's regularity lemma is a statement about the existence of finding a regular partitioning in arbitrary graphs, with a number K of partitions effectively independent of the size N of the graph.
\begin{definition}[Szemer\'edi's regularity lemma]\label{SRL}\cite{Diestel_2012}
For every $\epsilon>0$ and every integer $K_{min}\geq 1$, there exists an integer $K_{max}$ such that every graph on $N\geq K_{min}$ nodes admits an $\epsilon$-regular graph partitioning in $K$ partitions, with $K_{min}\leq K\leq K_{max}$.
\end{definition}
The proof of SRL can be found in Diestel \cite{Diestel_2012}. We exemplify the lemma: if we take a certain $\epsilon$ and choose $K_{min}=10$, then SRL states that there is an integer $K_{max}$, such that for any graph with $N>10$ nodes there exists an $\epsilon$-regular partitioning of $10\leq K\leq K_{max}$ partitions. While for $N\leq K_{max}$, the existence of an $\epsilon$-regular partitioning automatically holds by choosing the $K=N$ partitioning, the result becomes stronger for $N>K_{max}$. For very large graphs, i.e. $N\gg K_{max}\geq K$, it is still always possible to have an $\epsilon$-regular $K$-partitioning. 
\\
An interesting interpretation of SRL is given by Tao \cite{Tao_2005} who states that, roughly speaking: ``SRL can be viewed as a structure theorem for large dense graphs, approximating such graphs to any specified accuracy by objects whose complexity is bounded independently of the number of nodes in the original graph". Applied to UMFF, this means that, for any large dense graph and any desired accuracy $\epsilon$, there exists a partitioning in $K\ll N$ partitions, such that the topological approximation of UMFF between most $(k,m)$ partition pairs ($k\neq m$) is $\epsilon$-accurate, in the sense that $(k,m)$ are $\epsilon$-regular partition pairs. While a regular graph partitioning does not imply any regularity conditions on the within-partition links, Diestel \cite{Diestel_2012} mentions that by choosing $K_{min}$ large ``we may increase the proportion of links running between different partition sets (rather than inside one), i.e. the proportion of links that are subject to the regularity assertion". In other words, if we take $K_{min}$ large enough for a given $\epsilon$, then most links will be \emph{between} partitions (rather than \emph{within}) and will thus satisfy the regularity conditions.
%------------------------------------------
\subsubsection*{Implications of SRL for UMFF}
We believe that SRL can be translated to a statement about the scaling behavior of the SIS process on large graphs. We will describe the conceptual idea here, realizing that a more rigorous investigation would be necessary to proof any of the claims.
\\
Since the regularity inequality \eqref{eq_e-regular} can be rewritten as \eqref{eq_e-regular rewritten}, which has the same form as the isoperimetric inequality, the $\epsilon$-regularity of a partition pair also implies that the UMFF topological approximation \eqref{approx_topological approximation} has an $\epsilon$-bounded approximation error (for subsets of sufficiently large size). For an $\epsilon$-regular graph partitioning with $K+1$ pairs, this isoperimetric interpretation then means that for most of the partition pairs ($\geq\epsilon K^2$) the UMFF topological approximation error is $\epsilon$-bounded. Finally, SRL indicates that for any chosen accuracy $\epsilon$ and sufficiently large minimum number of partitions $K_{min}$, an integer $K_{max}$ exists such that for any graph on $N\geq K_{min}$ nodes, a partitioning can be found with $K_{min}\leq K\leq K_{max}$ partitions, such that most links are between partitions and most of the partition pairs have $\epsilon$-bounded approximation errors. Applied to UMFF, this means that \emph{for large graphs on $N$ nodes, a partitioning in $K_{min}<K\ll N$ partitions can always be found such that the UMFF topological approximation between most partition pairs is bounded by a chosen $\epsilon$, where choosing a large enough $K_{min}$ results in most links being between partitions} (by Diestel's argument). The UMFF approximation being $\epsilon$-bounded on a large graph implies that the dynamics of the SIS process on that graph can approximately be described by the dynamics on a much smaller, weighted graph of dimension $K\ll N$.
\\
\emph{Remark:} The regularity of SRL only holds for subsets of size $N_x\geq \epsilon N_k$, where $N_k\approx \frac{N}{K}$. Hence, the regularity weakens for growing $N$, because it no longer holds for cut-sets between small subsets. The consequence for UMFF is that the regularity, and thus the boundedness of the topological approximation error, only holds, if a sufficiently large fraction of nodes is infected in both partitions, i.e. $N_x$ infected nodes in $\mathcal{N}_k$ and $N_y$ in $\mathcal{N}_m$ for any $(k,m)$. Thus, the dynamics are well approximated by lower-dimensional dynamics, only for disease states where enough nodes are infected between any pair of partitions.

%% file: Sections/Related_work.tex
\section{Related work}
\label{S_Related work}
%-----------------------------
\subsubsection*{NIMFA on graphs with an equitable partitioning}
Bonaccorsi et al. \cite{Bonaccorsi_2015} study the NIMFA equations on graphs with an equitable partitioning. A partitioning $\pi$ is equitable if the subgraph between any two (possibly the same) partitions, is bi-regular (regular). If a graph has such an equitable partitioning, and the initial infection probability is the same for all nodes within one partition, then the NIMFA equations for the SIS process on that graph can be exactly described by $K$ rather than $N$ equations \cite{Bonaccorsi_2015}. This result follows from the observation that equality in the UMFF topological approximation \eqref{approx_topological approximation} holds, i.e.
$$
(u-w)^TA^{(km)}w = (\tilde{u}-\tilde{w})^T\widetilde{A}^{(km)}\tilde{w} = \tilde{a}_{km}(N_k-\tilde{w}_k)\tilde{w}_m
$$
when $A^{(km)}$ is bi-regular, and that
$$
\Pr[W(0)=w] = \left\vert{\mathcal{W}^k_{\tilde{w}_k}\cap\mathcal{W}^m_{\tilde{w}_m}}\right\vert^{-1}\Pr\left[\widetilde{W}_k(0)=\tilde{w}_k,\widetilde{W}_m(0)=\tilde{w}_m\right] \quad \forall w\in\mathcal{W}^k_{\tilde{w}_k}\cap\mathcal{W}^m_{\tilde{w}_m}
$$
holds, when nodes from the same partition have equal initial infection probabilities. Hence, the main point of \cite{Bonaccorsi_2015} is that for this specific type of graph and initial condition, the number of infective links between any two partitions only depends on the number of infected nodes in those partitions, which enables a lower-dimensional description of the SIS process (within the NIMFA approximation). This result is based on similar ideas as the UMFF framework, but from a very different perspective: UMFF describes how the topological approximation \eqref{approx_topological approximation} applied to \emph{any graph}, followed by a moment-closure approximation \eqref{approx_moment-closure approximation}, results in a lower-dimensional \emph{approximate} description of the SIS process.
%---------------------------------------------------
\subsubsection*{Approximating the number of infective links in SIS}
A central concept of UMFF is the description of the topological approximation \eqref{approx_topological approximation} from the perspective of the isoperimetric problem. This approach of approximating the SIS process by approximating the number of infective links has appeared before.
\\
Ganesh et al. \cite{Ganesh_2005} find an upper-bound for the epidemic threshold, by relating the infection terms in the SIS process to the isoperimetric problem. The isoperimetric or Cheeger constant \cite{Van_Mieghem_GS} of a graph with adjacency matrix $A$ is defined as:
$$
\eta_{c}(A) = \min_{w\in\lbrace{0,1}\rbrace^N}\frac{(u-w)^TAw}{w^Tw}
$$
which leads to a lower-bound for the number of infective links as:
\begin{equation}\label{eq_ganesh}
(u-w)^TAw \geq \eta_c(A)\tilde{w}
\end{equation}
for any $w\in\lbrace{0,1}\rbrace^N$ and where $\tilde{w}=w^Tw$ is the number of infected nodes. By assuming equality in \eqref{eq_ganesh}, the SIS process is approximated by a linear death-birth process, from which an approximate epidemic threshold is derived in \cite{Ganesh_2005}.
\\
\\
Van Mieghem \cite{Van_Mieghem_2016}, \cite{Van_Mieghem_2016B} also approximated the SIS process by approximating the size of the cut-set. Rather than relying on the isoperimetric problem, the most dominant terms in the spectral decomposition of the quadratic form $w^TQw$, which equals the number of infective links, approximate the cut-set. Specifically, the approximation
$$
(u-w)^TAw \approx \frac{\mu_{N-1}}{N}\tilde{w}(N-\tilde{w}) 
$$
is made. If this approximation error can be bounded by a constant $\theta\in\mathbb{R}$, i.e.
\begin{equation}\label{eq_approx bounds}
\left\vert (u-w)^TAw - \frac{\mu_{N-1}}{N}\tilde{w}(N-\tilde{w}) \right\vert \leq \theta
\end{equation}
then the exact equation for the expected number of infected nodes can be bounded as
\begin{equation} \label{eq_process bound}
\mathbf{E}[\widetilde{W}_{-\theta}(t)]\leq \mathbf{E}[\widetilde{W}_{\text{exact}}(t)] \leq \mathbf{E}[\widetilde{W}_{+\theta}(t)]
\end{equation}
where the bounds follow from the differential equations:
\begin{equation}\label{eq_bounds}
\begin{dcases}
\frac{d\mathbf{E}[\widetilde{W}_{+\theta}(t)]}{dt} = -\delta\mathbf{E}[\widetilde{W}_{+\theta}] + \beta\frac{\mu_{N-1}}{N}\mathbf{E}[\widetilde{W}_{+\theta}](N-\mathbf{E}[\widetilde{W}_{+\theta}]) + \theta \\
\frac{d\mathbf{E}[\widetilde{W}_{-\theta}(t)]}{dt} = -\delta\mathbf{E}[\widetilde{W}_{-\theta}] + \beta\frac{\mu_{N-1}}{N}\mathbf{E}[\widetilde{W}_{-\theta}](N-\mathbf{E}[\widetilde{W}_{-\theta}]) - \theta
\end{dcases}
\end{equation}
which are Riccati differential equations, whose analytic solution are known and have a hyperbolic-tangent form \cite{Van_Mieghem_2016}. In other words, the method of \cite{Van_Mieghem_2016} and \cite{Van_Mieghem_2016B} gives bounds on the exact expected number of infected nodes $\mathbf{E}[\widetilde{w}_{\text{exact}}(t)]$, if a constant bound $\theta$ on the approximation error \eqref{eq_approx bounds} is known. 
\\
By filling in $c=\mu_{N-1}$ in Lemma \ref{L_general isoperimetric inequality} from Appendix \ref{A_proof isoperimetric}, we can show that $\theta \leq \frac{N(\mu_1-\mu_{N-1})}{4}=\theta^{\star}$. Although not a tight bound, filling in $\theta=\theta^{\star}$ in equations \eqref{eq_bounds} gives:
$$
\mathbf{E}[\widetilde{W}(t)_{-\theta^{\star}}] \leq \mathbf{E}[\widetilde{W}_{\text{exact}}(t)] \leq\mathbf{E}[\widetilde{W}(t)_{+\theta^{\star}}]
$$
which is a new result based on the spectral decomposition methodology of \cite{Van_Mieghem_2016} and \cite{Van_Mieghem_2016B}.

%% file: Sections/Conclusion.tex
\section{Summary}
\label{S_conclusion}
We have introduced a novel approximation framework for the description of the Markovian SIS process on complex networks. The main features of this Universal Mean-Field Framework (UMFF) are: 
\begin{itemize}
\item UMFF unifies and generalizes a number of existing mean-field methods for approximating SIS epidemics on complex networks. In particular, two widely-used techniques, the N-Intertwined Mean-Field Approximation \cite{Van_Mieghem_2009} and the Heterogeneous Mean-Field method \cite{Pastor-Satorras_2001} are shown to be contained by UMFF.
\item The accuracy of UMFF and of all its the contained methods can be assessed based on the isoperimetric analogy:
\begin{align*}
\text{infected nodes} &\leftrightarrow \text{graph volume} \\ 
\text{infective links} &\leftrightarrow \text{graph surface}
\end{align*}
which provides bounds on the error of the UMFF topological approximation.
\item UMFF leads to a conceptual description of the scaling behavior of SIS epidemics on large graphs. Since the UMFF accuracy is related to the notion of regularity on which Szemer\'edi's regularity lemma (SRL) is based, we can translate the statements of SRL about the structural regularity of large graphs to statements about the possibility to accurately approximate SIS dynamics on large graphs by a lower-dimensional description.
\end{itemize}
\subsubsection*{Future directions}
By providing a universal description of mean-field approximation techniques for the SIS process, UMFF offers a framework, in which the existing techniques can be compared and which enables their respective accuracy to be assessed. In principle, UMFF could prescribe which existing (or new) mean-field method is more suitable, for a certain graph specification and for a specific SIS process parameter of interest.
\\
While derived specifically for SIS epidemics, the UMFF approach is applicable to more general epidemic models. Sahneh et al. \cite{Sahneh_2013} for instance, describe the Generalized Epidemic Mean-Field model (GEMF), which is a generalization of the NIMFA approach to epidemic models with any number of compartments, and with a general transition structure between different compartments. The global dynamics of this general epidemic model follow from \emph{node-based} compartmental transitions and \emph{edge-based} compartmental transitions, which translates to \emph{volume-based} transitions and \emph{surface-based} transitions in context of the isoperimetric problem. Hence, by exploiting the same problem structure and the isoperimetric analogy, UMFF could generalize GEMF in a similar vein as UMFF generalizes NIMFA for the SIS compartmental process.
\\
The general partitioning feature of UMFF also creates the possibility to develop new approximation techniques for the SIS process. Specifically, if nodes can be grouped in partitions based on some parameter such that similarity in that parameter corresponds to similarity in connectivity, then UMFF is expected to yield a good approximation of the SIS process. For instance, the embedding of graphs in metric spaces is studied in \cite{Serrano_2008} and \cite{Krioukov_2010}. Nodes are considered to be embedded in a metric space with linking probabilities between node pairs dependent on the distance between them. Similarity in spatial coordinates (i.e. a small distance) between a pair of nodes means that their distance to other nodes is also similar. Hence, for such graph models, spatial closeness of nodes seems to provide a good partitioning criterion for UMFF, and the coarse-graining of the infection state would then correspond to the intuitively attractive notion of spatial coarse-graining.
\\
Furthermore, the observation that both the exact and approximate Markovian SIS processes are equivalent to a higher-dimensional quadratic death-birth process opens up new perspectives on modeling the SIS process. Some questions about the epidemic process have tractable solutions if properly formulated in terms of death-birth processes. Ganesh et al. \cite{Ganesh_2005} for instance, characterized the disease die-out probability of the SIS process, based on the gambler's ruin problem \cite{Van_Mieghem_PA} of a death-birth process. Conversely, the knowledge about the epidemic process might provide valuable insights in the quadratic death-birth process, whose exact solution is still an open problem \cite{Van_Mieghem_2017}.

%% file: Appendix/A_Kolmogorov_equations.tex
\section{Derivation of exact SIS equations for $\widetilde{W}$ and $\mathbf{E}[\widetilde{W}]$}
\label{A_exact equations}
\subsection{The Kolmogorov equations for Markov Chains: brief reminder}
\label{ASS_kolmogorov equations}
As a background for the further derivation of the UMFF equations \eqref{eq_UMFF}, we start with a toy example to illustrate how the Kolmogorov equations are found for a Markov Chain. Further details can be found in \cite{Van_Mieghem_PA}. Consider the 3-state Markov chain in $W(t)$ below:
\\
\input{./Figures/mc.tikz}
\\
The Markov chain has three states: $w_1$,$w_2$ and $w_3$, with state probabilities $\Pr[W(t)={w}_i]$ and transition rates $r_{ij}$, for $1\leq i \neq j\leq 3$. By the subscript ``$ij$" in the rates $r_{ij}$, we denote the transition from state $i$ to state $j$, i.e. $i\rightarrow j$. As mentioned in Section \ref{S_background}, we assume that the transition processes are independent Poisson processes with exponentially distributed inter-event times, for example for the transition $r_{12}$ this yields:
$$
\Pr[W(t+h)=w_2\vert W(t)=w_1] = r_{12} e^{-r_{12}h}
$$
For $h\rightarrow 0$, this transition leads to:
$$
\begin{cases}
\frac{d\Pr[W(t)=w_2]}{dt} = r_{12} \Pr[W(t)=w_1] \\
\frac{d\Pr[W(t)=w_1]}{dt} = -r_{12} \Pr[W(t)=w_1] 
\end{cases}
$$
Combining all transitions then leads to the Kolmogorov equations:
$$
\begin{cases}
\frac{d\Pr[W(t)=w_1]}{dt} = -r_{12} \Pr[W(t)=w_1] +r_{21}\Pr[W(t)=w_2] \\
\frac{d\Pr[W(t)=w_2]}{dt} = r_{12}\Pr[W(t)=w_1] - (r_{23}+r_{21})\Pr[W(t)=w_2] + r_{32}\Pr[W(t)=w_3] \\
\frac{d\Pr[W(t)=w_3]}{dt} = r_{23}\Pr[W(t)=w_2] - r_{32}\Pr[W(t)=w_3]
\end{cases}
$$
Hence, by identifying the state transitions and according rates, one obtains the Kolmogorov equations of a Markov Chain, which completely characterize the dynamics of the process for a given initial distribution $\Pr[W(0)=w_i]$, for each possible state $w$.
%--------------------------------------------------------------------
\subsection{State probability $\Pr[\widetilde{W}(t)=\tilde{w}]$}
\label{ASS_state probability}
As described in Sections \ref{S_definition of GHMF} and \ref{S_derivation of GHMF}, the reduced-state vector $\tilde{w}$ is introduced to compactly describe the disease state and to reduce the complexity of the SIS process description. Instead of describing the state of each node separately, the reduced-state vector $\tilde{w}=\left(\tilde{w}_1,\tilde{w}_2,\dots,\tilde{w}_K \right)$ captures the number of infected nodes in each partition, by the relation $\tilde{w}_k=s_k^Tw$. By $\mathcal{W}^k_{x}=\left\lbrace w\in\lbrace 0,1\rbrace^N\vert s_k^Tw = x \right\rbrace$ we denote the set of all full-state vectors $w$ with $x$ nodes infected in partition $k$ (and with any possible number of nodes nodes in the other partitions $m\neq k$). Each full-state vector $w\in\bigcap_{k=1}^K\mathcal{W}^k_{\tilde{w}_k}$ then corresponds to the reduced-state vector $\tilde{w}$, since each set $\mathcal{W}^k_{\tilde{w}_k}$ constrains the number of infected nodes in a specific partition $k$. Based on this notation, we can represent the coarse-graining of the full states to the reduced states as:
$$
\bigcap_{k=1}^K\mathcal{W}^k_{\tilde{w}_k} \xrightarrow{\text{group by partitioning } \pi} \tilde{w}
$$
The full-state and reduced-state probabilities are then related as:
\begin{equation}\label{eq_prob relation}
\Pr[\widetilde{W}=\tilde{w}] = \sum_{w\in\bigcap_{k=1}^K\mathcal{W}^k_{\tilde{w}_k}}\Pr[W=w]
\end{equation}
and similarly, the rates are related as:
\begin{equation}\label{eq_rate relation}
r_{\tilde{w}_k(\tilde{w}_k\pm \tilde{e}_k)}\Pr[\widetilde{W}_k=\tilde{w}_k] = \sum_{w\in\bigcap_{i=1}^K\mathcal{W}^i_{\tilde{w}_i}}\sum_{j\in\mathcal{N}_k} r_{w(w+e_j)}\Pr[W=w]
\end{equation}
\\
More can be said about the reduced-state transition structure: firstly, the entries $\tilde{w}_k$ represent the number of infected nodes in partition $k$, from which it follows that 
$$
\tilde{w}\in\lbrace{0,1,\dots,N_1}\rbrace\times\lbrace{0,1,\dots,N_2}\rbrace\times\dots\times\lbrace{0,1,\dots,N_K}\rbrace,
$$
and secondly, since a state transition in the Markovian SIS process reflects a single infection or curing event, the possible transitions between reduced states are of the form 
$$
\tilde{w} \rightarrow \tilde{w} \pm \tilde{e}_k
$$
Hence, the reduced states and their transitions constitute an $(N_1+1)\times(N_2+1)\times\dots\times(N_K+1)$ lattice. This structure can be represented compactly by the chain below, which depicts one specific ``direction" in the lattice, corresponding to one partition $k$.
\\
\input{./Figures/mc_2.tikz}
\\
However, since the transition rates between reduced states depend on the full states \eqref{eq_rate relation}, the transitions at the reduced-state level do not describe a Markov chain. Nonetheless, it is still possible to write the exact, but not-closed differential equations for the reduced-state probabilities by grouping the Kolmogorov equations according to the partitions:
$$
\frac{d\Pr[\widetilde{W}=\tilde{w}]}{dt} = \sum_{w\in\bigcap_{k=1}^K\mathcal{W}^k_{\tilde{w}_k}}\frac{d\Pr[W=w]}{dt}
$$
Considering the transitions within the partitions separately enables the Kolmogorov equations at the reduced-state level to be written as:
\begin{equation} \label{A_Kolmogorov}
\begin{split}
\frac{d\Pr[\widetilde{W}(t)=\tilde{w}]}{dt} = \sum_{k=1}^K\Big( &-r_{\tilde{w}(\tilde{w}-\tilde{e}_k)}\Pr[\widetilde{W}=\tilde{w}] +r_{(\tilde{w}+\tilde{e}_k)\tilde{w}}\Pr[\widetilde{W}=\tilde{w}+\tilde{e}_k] \\
&-r_{\tilde{w}(\tilde{w}+\tilde{e}_k)}\Pr[\widetilde{W}=\tilde{w}] +r_{(\tilde{w}-\tilde{e}_k)\tilde{w}}\Pr[\widetilde{W}=\tilde{w}-\tilde{e}_k] \Big)
\end{split}
\end{equation}
The transition rates at the reduced-state level are derived below.
%--------------------------------------------------------
\subsubsection*{Transition rates $r_{\tilde{w}(\tilde{w}-\tilde{e}_k)}$ and $r_{(\tilde{w}+\tilde{e}_k)\tilde{w}}$: node healing in partition $k$.}
By the grouping relation \eqref{eq_rate relation} between the full states and the reduced states, the reduced-state transition rates are given by:
\begin{equation}\label{eq_healing}
r_{\tilde{w}(\tilde{w}-\tilde{e}_k)}\Pr[\widetilde{W}=\tilde{w}] = \sum_{w\in\bigcap_{i=1}^K\mathcal{W}^i_{\tilde{w}_i}}\sum_{j\in\mathcal{N}_k}r_{w(w-e_j)}\Pr[W=w]
\end{equation}
The transition rate $r_{w(w-e_j)}$ in equation \eqref{eq_healing} corresponds to node $j$ healing in state $w$, i.e. the transition $W_j=1\rightarrow W_j=0$. The healing rate in UMFF is $\delta$ for any node, hence the transition rate equals
$$
r_{w(w-e_j)} = \delta w_j
$$
for any full-state vector $w$ and node $j$.
The sum of the healing rates for all nodes in a partition $k$ is then:
\begin{equation}\label{eq_healing partition}
\sum_{j\in\mathcal{N}_k} r_{w(w-e_j)} = \delta s_k^Tw = \delta\tilde{w}_k
\end{equation}
Substituting \eqref{eq_healing partition} in the rate equation \eqref{eq_healing} and invoking \eqref{eq_prob relation} then yields
\begin{equation}\label{eq_rate 1}
r_{\tilde{w}(\tilde{w}-\tilde{e}_k)}\Pr[\widetilde{W}=\tilde{w}] = \delta\tilde{w}_k\Pr[\widetilde{W}=\tilde{w}]
\end{equation}
for the reduced-state transition rate corresponding to a node healing in partition $k$, in state $\tilde{w}$. A similar derivation yields 
\begin{equation}\label{eq_rate 2}
r_{(\tilde{w}+\tilde{e}_k)\tilde{w}}\Pr[\widetilde{W}=\tilde{w}+\tilde{e}_k] = \delta(\tilde{w}_k+1)\Pr[\widetilde{W}=\tilde{w}+\tilde{e}_k]
\end{equation}
for the reduced-state transition rate corresponding to a node healing in partition $k$, in state $\tilde{w}+\tilde{e}_k$.
%---------------------------------------------------
\subsubsection*{Transition rates $r_{\tilde{w}(\tilde{w}+\tilde{e}_k)}$ and $r_{(\tilde{w}-\tilde{e}_k)\tilde{w}}$: a node in partition $k$ is infected.}
By the grouping relation \eqref{eq_rate relation} between the full states and the reduced states, the reduced-state transition rates are given by equation:
\begin{equation}\label{eq_infection}
r_{\tilde{w}(\tilde{w}+\tilde{e}_k)}\Pr[\widetilde{W}=\tilde{w}] = \sum_{w\in\bigcap_{i=1}^K\mathcal{W}^i_{\tilde{w}_i}}\sum_{j\in\mathcal{N}_k}r_{w(w+e_j)}\Pr[W=w]
\end{equation}
The transition rate $r_{w(w+e_j)}$ in equation \eqref{eq_infection} corresponds to node $j$ becoming infected in state $w$, i.e. the transition $W_j=0\rightarrow W_j=1$. Since 
$$
e_j^TAw = \sum_{i=1}^Na_{ij}w_i
$$
is the number of infected neighbors of node $j$, and since each infected neighbor infects node $j$ at a rate $\beta$ if $w_j=0$, the full-state transition rate
$$
r_{w(w+e_j)} = \beta(1-w_j)e_j^TAw
$$
is found. The sum of infection rates for all nodes in partition $k$ is then:
\begin{equation}\label{eq_infection partition}
\sum_{j\in\mathcal{N}_k} = \beta\sum_{m=1}^K(u-w)^TA^{(km)}w
\end{equation}
where the sum over partitions $1\leq m\leq K$ is introduced such that the block-matrix $A^{(km)}$, which naturally reflects the partition structure, can be used. Filling \eqref{eq_infection partition} into the rate equation \eqref{eq_infection} then yields:
\begin{equation}\label{eq_rate 3}
r_{\tilde{w}(\tilde{w}+\tilde{e}_k)}\Pr[\widetilde{W}=\tilde{w}] = \beta\sum_{m=1}^K\sum_{w\in\mathcal{W}^{k}_{\tilde{w}_k}\cap\mathcal{W}^m_{\tilde{w}_m}} (u-w)^TA^{(km)}w\Pr[W=w]
\end{equation}
for the reduced-state transition rate corresponding to a node becoming infected in partition $k$, in state $\tilde{w}$. A similar derivation yields 
\begin{equation}\label{eq_rate 4}
r_{(\tilde{w}-\tilde{e}_k)\tilde{w}}\Pr[\widetilde{W}=\tilde{w}-\tilde{e}_k] = \beta\sum_{m=1}^K\sum_{w\in\mathcal{W}^{k}_{(\tilde{w}_k-1)}\cap\mathcal{W}^m_{\tilde{w}_m}} (u-w)^TA^{(km)}w\Pr[W=w]
\end{equation}
for the reduced-state transition rate corresponding to a node becoming infected in partition $k$, in state $\tilde{w}-\tilde{e}_k$.
%-----------------
\subsubsection*{Resulting reduced-state equations}
Introducing the rates \eqref{eq_rate 1},\eqref{eq_rate 2},\eqref{eq_rate 3} and \eqref{eq_rate 3} the Kolmogorov equations \eqref{A_Kolmogorov} establishes equation \eqref{eq_state prob exact} in Section \ref{S_derivation of GHMF}.
%--------------------------------------------------------
\subsection{Expected number of infected nodes $\mathbf{E}[\widetilde{W}_k]$}
\label{ASS_expected number}
The equations for the expected number of infected nodes $\mathbf{E}[\widetilde{W}_k]$ can be derived from the reduced-state probability equations \eqref{eq_state prob exact}, based on the definition of expectation and the law of total probability. For any partition $k$, we can write the expected number of infected nodes as:
\begin{equation}\label{eq_expectation definition}
\mathbf{E}[\widetilde{W}_k] = \sum_{\tilde{w}_k=0}^{N_k}\tilde{w}_k\Pr[\widetilde{W}_k=\tilde{w}_k]
\end{equation}
Since by the law of total probability, the marginal probability can be written as:
$$
\Pr[\widetilde{W}_k=\tilde{w}_k] = \underbrace{\sum_{\tilde{w}_1=0}^{N_1}\dots\sum_{\tilde{w}_l=0}^{N_l}\dots\sum_{\tilde{w}_K=0}^{N_K}}_{\forall l\neq k}\Pr[\widetilde{W}=\tilde{w}]
$$
such that \eqref{eq_expectation definition} equals
\begin{equation}\label{eq_expectation definition full}
\mathbf{E}[\widetilde{W}_k] = \underbrace{\sum_{\tilde{w}_1=0}^{N_1}\dots\sum_{\tilde{w}_l=0}^{N_l}\dots\sum_{\tilde{w}_K=0}^{N_K}}_{\forall l} \tilde{w}_k\Pr[\widetilde{W}=\tilde{w}]
\end{equation}
Differentiation with respect to time of equation \eqref{eq_expectation definition full} then yields:
\begin{equation}\label{eq_expectation definition final}
\frac{d\mathbf{E}[\widetilde{W}_k]}{dt} = \underbrace{\sum_{\tilde{w}_1=0}^{N_1}\dots\sum_{\tilde{w}_l=0}^{N_l}\dots\sum_{\tilde{w}_K=0}^{N_K}}_{\forall l} \tilde{w}_k\frac{d\Pr[\widetilde{W}=\tilde{w}]}{dt}
\end{equation}
After substitution of $\frac{d\Pr[\widetilde{W}=\tilde{w}]}{dt}$ from equation \eqref{eq_state prob exact}, we arrive at equation \eqref{eq_prevalence exact} in Section \ref{S_derivation of GHMF}.

%% file: Figures/mc.tikz
%Source: http://steventhornton.ca/markov-chains-in-latex/
\begin{figure}[h!]
\centering
\begin{tikzpicture}
        % Draw the states
        \node[state]             (w1) {$w_1$};
        \node[state, right=of w1] (w2) {$w_2$};
        \node[state, right=of w2] (w3) {$w_3$};
 
        % Connect the states with arrows
        \draw[every loop]
            (w1) edge[bend right, auto=left] node {$r_{12}$} (w2)
            (w2) edge[bend right, auto=left] node {$r_{23}$} (w3)
            (w3) edge[bend right, auto=right] node {$r_{32}$} (w2)
            (w2) edge[bend right, auto=right] node {$r_{21}$} (w1);
\end{tikzpicture}
\end{figure}

%% file: Figures/mc_2.tikz
%Source: http://steventhornton.ca/markov-chains-in-latex/
\begin{figure}[h!]
\centering
\begin{tikzpicture}
        % Draw the states
        \node[state]             (w1) {$\tilde{w}-\tilde{e}_k$};
        \node[state, right=2.5cm of w1] (w2) {$~~~\tilde{w}~~~$};
        \node[state, right=2.5cm of w2] (w3) {$\tilde{w}+\tilde{e}_k$};
        \node[draw=none, left=of w1] (dleft) {$\dots$};
        \node[draw=none, right=of w3] (dright) {$\dots$};
 
        % Connect the states with arrows
        \draw[every loop]
            (w1) edge[bend right, auto=left] node {$r_{(\tilde{w}-\tilde{e}_k)\tilde{w}}$} (w2)
            (w2) edge[bend right, auto=left] node {$r_{\tilde{w}(\tilde{w}+\tilde{e}_k)}$} (w3)
            (w3) edge[bend right, auto=right] node {$r_{(\tilde{w}+\tilde{e}_k)\tilde{w}}$} (w2)
            (w2) edge[bend right, auto=right] node {$r_{\tilde{w}(\tilde{w}-\tilde{e}_k)}$} (w1)
                   
            %(w2) edge[bend right, auto=left] node {$r_{24}$} (w4)
            %(w4) edge[bend right, auto=right] node {$r_{42}$} (w2)
            %(w2) edge[bend right, auto=left] node {$r_{25}$} (w5)
            %(w5) edge[bend right, auto=right] node {$r_{52}$} (w2)
            
            %(w4) edge[bend right, auto=left] node {} (dabove)
            %(dabove) edge[bend right, auto=right] node {} (w4)
            %(w5) edge[bend right, auto=left] node {} (dbelow)
            %(dbelow) edge[bend right, auto=right] node {} (w5)
            
            (w1) edge[bend right, auto=right] node {} (dleft)
            (dleft) edge[bend right, auto=left] node {} (w1)
            (dright) edge[bend right, auto=right] node {} (w3)
            (w3) edge[bend right, auto=left] node {} (dright)
            ;
\end{tikzpicture}
\end{figure}

%% file: Appendix/A_Higher-order_UMFF.tex
\section{Higher-order UMFF}
\label{A_higher-order UMFF}
The UMFF equations can be extended to higher-order moments, in order to better capture the dynamic correlations of the SIS process.
\\
For the case of $K=N$ partitions, Cator et al. \cite{Cator_2012} and Mata et al. \cite{Mata_2013} have described how the NIMFA \cite{Van_Mieghem_2009} and Quenched Mean-Field (QMF) \cite{Castellano_2010} equations can be extended to $n$'th-order moments:
$$
\mathbf{E}[W_i]\rightarrow \mathbf{E}[W_i],\mathbf{E}[W_iW_j],\dots,\mathbf{E}[\underbrace{W_iW_j\dots W_l}_{n}]
$$
based on the exact SIS dynamics. In order to have a closed set of equations for order $n$, the $(n+1)$'th-order moments must be approximated by lower-order moments, i.e. an approximation of the form:
$$
\mathbf{E}[\underbrace{W_iW_j\dots W_l}_{n+1}] \approx f\left(\mathbf{E}[\underbrace{W_iW_j\dots W_l}_{\forall m\leq n}]\right)
$$
where different choices for the moment-closure approximation $f$ are given in \cite{Cator_2012} and \cite{Mata_2013}. Similarly, we can define the higher-order Universal Mean-Field Framework as:
\begin{definition}[Higher-order UMFF]
Consider a graph $\mathcal{G}(\mathcal{N},\mathcal{L})$, an epidemic process with rates $(\beta,\delta)$ and a graph partitioning $\pi$. For any integer $n\leq K$, the $n$'th-order UMFF equations are given by:
\begin{equation} \label{eq_higher-order UMFF def}
\frac{d\mathbf{E}\left[\prod_{k=1}^K\widetilde{W}_k^{p_k}\right]}{dt} = \underbrace{\sum_{\tilde{w}_1=0}^{N_1} \dots \sum_{\tilde{w}_k=0}^{N_k} \dots \sum_{\tilde{w}_K=0}^{N_K}}_{\forall k}\prod_{k=1}^K\tilde{w}_k^{p_k}\frac{d\Pr[\widetilde{W}=\tilde{w}]}{dt}
\end{equation}
for all vectors $p\in\left\lbrace{p\in\mathbb{N}^K\vert 0\leq p_k\leq N_k,\forall k\text{ and }u^Tp\leq n}\right\rbrace$ and with $\frac{d\Pr[\widetilde{W}=\tilde{w}]}{dt}$ given by equation \eqref{eq_state prob topological approximation 1}.
The $(n+1)$'th-order moments appearing in the higher-order UMFF equations are approximated by:
\begin{equation}\label{eq_UMFF moment-closure}
\mathbf{E}\left[\prod_{k=1}^K\widetilde{W}_k^{p_k}\right] \approx f\left(
\left\lbrace{
\mathbf{E}\left[\prod_{k=1}^K\widetilde{W}_k^{q_k}\right]
}\right\rbrace_{\forall q\in\mathcal{Q}}
\right)
\end{equation}
for all vectors $p\in\left\lbrace{p\in\mathbb{N}^K\vert 0\leq p_k\leq N_k,\forall k\text{ and }u^Tp=n+1}\right\rbrace$ and with \\ $\mathcal{Q}=\left\lbrace{q\in\mathbb{N}^K\vert 0\leq q_k\leq p_k,\forall k\text{ and }u^Tq\leq n}\right\rbrace$. The function $f$ represents a generic moment-closure approximation.
\end{definition}
\textit{Remark (1):} The higher-order UMFF equations are found from the definition of expectation and the law of total probability, similar to the derivation of the first-order moments in Appendix \ref{ASS_expected number}.
\\
\textit{Remark (2):} For a certain partition $k$, only the moments $\mathbf{E}\left[\dots\widetilde{W}_k^{p_k}\dots\right]$ for values $p_k\in\lbrace{1,\dots,N_k}\rbrace$ are considered. Since $\tilde{w}_k\in\lbrace{0,1,\dots,N_k}\rbrace$ has $(N_k+1)$ possible values, the probability distribution $\Pr[\widetilde{W}_k=\tilde{w}_k]$ is fully determined by the first $N_k$ moments. Hence the set $\left\lbrace{p\in\mathbb{N}^K\vert 0\leq p_k\leq N_k,\forall k\text{ and }u^Tp\leq n}\right\rbrace$ represents the set of powers of all $n$'th-order moments.

%% file: Appendix/A_Proof_of_isoperimetric_inequalities.tex
\section{Proof of isoperimetric inequalities}
\label{A_proof isoperimetric}
In this section, we prove the isoperimetric inequalities \eqref{topological approx general graph} and \eqref{topological approx bi-regular graph} of Theorem \ref{topological approx theorem} and Theorem \ref{bi-regular approx theorem}. We start by introducing some definitions and notations, based on the work of Haemers \cite{Haemers_1995}. We then state and prove Lemma \ref{L_general isoperimetric inequality}, from which  Theorem \ref{topological approx theorem} follows. Finally, we prove Theorem \ref{bi-regular approx theorem} based on the specific structure of bi-regular graphs.
%-------------------------------------------------
\subsection{Interlacing and quotient matrices}
The following definitions are given in \cite{Haemers_1995} and \cite{Van_Mieghem_GS}:
\begin{definition}[Interlacing sequences]
Consider two sequences of real numbers: $\alpha_1\geq \alpha_2\geq\dots\geq \alpha_N$ and $\gamma_1\geq \gamma_2\geq\dots\geq \gamma_K$ with $K\leq N$. The second sequence is said to interlace the first whenever
\begin{equation}\nonumber
\alpha_i \geq \gamma_i \geq \alpha_{N-K+i} \quad \text{for } i=1,\dots,K
\end{equation}
\end{definition}
\begin{definition}[Quotient matrix]
The quotient matrix $A^{(\pi)}$ of an adjacency matrix $A$ according to a partitioning $\pi$, is the matrix whose entries are the average row sums of the blocks of $A$. More precisely, $a^{(\pi)}_{km}$ is the entry in the quotient matrix according to the submatrix of $A$ between nodes of $\mathcal{N}_k$ and $\mathcal{N}_m$ with value
\begin{equation}
a^{(\pi)}_{km} = \frac{1}{N_k}s_k^TAs_m
\nonumber
\end{equation}
\end{definition}
These concepts can be combined by the interlacing theorem \cite{Haemers_1995}:
\begin{theorem}[Interlacing theorem] \label{theorem_quotient interlacing}
Suppose $A^{(\pi)}$ is the quotient matrix of a matrix $A$, then the eigenvalues of $A^{(\pi)}$ interlace the eigenvalues of $A$.
\end{theorem}
The interlacing theorem is crucial for the proof of the isoperimetric inequality as will become clear in the proof of Lemma \ref{L_general isoperimetric inequality}.
%----------------------------------------------
\subsection{General isoperimetric inequality}
We start by proving Lemma \ref{L_general isoperimetric inequality} below:
\begin{lemma} \label{L_general isoperimetric inequality}
Consider a graph $G(\mathcal{N},\mathcal{L})$ with $N$ nodes. For any $c\in\mathbb{R}$ and any pair of Bernoulli vectors $w_x,w_y\in\lbrace{0,1}\rbrace^N$, with $N_x=u^Tw_x$ and $N_y=u^Tw_y$ ones, respectively, and with $w_x^Tw_y=0$, the following inequality holds:
\begin{equation} \label{eq_lemma}
\left\vert w_x^TAw_y - \frac{c}{N}N_xN_y\right\vert \leq \frac{\theta}{N}\sqrt{N(N-N_x)N_y(N-N_y)}
\end{equation}
where $\vert c-\mu_i\vert \leq \theta$ for $1\leq i< N$ holds.
\end{lemma}
A first proof of Lemma \ref{L_general isoperimetric inequality} is given by Chung \cite{Chung_1996} in the context of isoperimetric inequalities and discrepancy inequalities on graphs. The proof is mainly based on algebraic manipulations of the term $w_x^TAw_y$ and the eigendecomposition of the Laplacian matrix $Q$. As mentioned in Section \ref{S_background}, the Laplacian $Q$ is a positive semidefinite matrix possessing the eigendecomposition:
$$
Q=ZMZ^T
$$
where $Z$ is the orthogonal eigen-matrix with eigenvectors $z_i$ as columns, and $M=\operatorname{diag}(\mu_1,\mu_2,\dots,\mu_N)$, the diagonal matrix containing the eigenvalues. These eigenvalues can be ordered as $0=\mu_N<\mu_{N-1}\leq\dots\leq\mu_1$, where the $0$ eigenvalue corresponds to the all-one eigenvector $z_N=\frac{u}{\sqrt{N}}$. Now, if we denote by $\widetilde{Z}$ the $N\times(N-1)$ matrix with $z_N$ removed, and by $\widetilde{M}$ the $(N-1)\times(N-1)$ diagonal matrix $\widetilde{M}=\operatorname{diag}(\mu_{1},\dots,\mu_{N-1})$, then we can also write:
$$
Q = \widetilde{Z}\widetilde{M}\widetilde{Z}^T
$$
If we further denote by $Q_K=NI-uu^T$ the Laplacian matrix of the complete graph, then we can write $\widetilde{Z}\widetilde{Z}^T=\frac{1}{N}Q_K$, which holds for $\widetilde{Z}$ of any Laplacian matrix.
%---
\\ \textbf{Proof A:} \\
We start by rewriting $w_x^TAw_y=w_x^T(\Delta-Q)w_y$. Due to the condition that $w_x^Tw_y=0$, we have $w_x^T\Delta w_y=0$ and thus $w_x^TAw_y=-w_x^TQw_y$. We then introduce the value $c\in\mathbb{R}$ as follows:
\begin{align*}
w_x^TAw_y &= -w_x^TQw_y +\frac{c}{N}w_x^TQ_Kw_y-\frac{c}{N}w_x^TQ_Kw_y \\
&= w_x^T(\frac{c}{N}Q_K-Q)w_y - \frac{c}{N}w_x^TQ_Kw_y
\end{align*}
Since $w_x^TQ_Kw_y=-N_xN_y$, and using the eigendecomposition of $Q$ and $Q_K$, we obtain:
$$
w_x^TAw_y = w_x^T\widetilde{Z}(cI-\widetilde{M})\widetilde{Z}^Tw_y + \frac{c}{N}N_xN_y
$$
or,
$$
w_x^TAw_y - \frac{c}{N}N_xN_y = w_x^T\widetilde{Z}(cI-\widetilde{M})\widetilde{Z}^Tw_y
$$
By introducing the variables $\alpha_i = w_x^Tz_i$ and $\beta_i=w_y^Tz_i$, we can write:
\begin{equation} \label{eqA_chung decomposed}
w_x^TAw_y - \frac{c}{N}N_xN_y = \sum_{i=1}^{N-1}\alpha_i\beta_i(c-\mu_i)
\end{equation}
We can upper-bound the right-hand side as $\left\vert\sum_{i=1}^{N-1}\alpha_i\beta_i(c-\mu_i)\right\vert \leq \theta\sum_{i=1}^{N-1}\vert \alpha_i\beta_i\vert$, where we introduce $\theta$ with $\vert c-\mu_i\vert\leq\theta,\forall i\neq N$ as an upper bound. Equation \eqref{eqA_chung decomposed} can then be written as:
$$
\left\vert w_x^TAw_y - \frac{c}{N}N_xN_y\right\vert \leq \theta\sum_{i=1}^{N-1}\vert \alpha_i\beta_i\vert
$$
Now, invoking the Cauchy-Schwartz inequality on the right-hand side of the equation and replacing $\alpha_i,\beta_i$ by their original values yields:
$$
\left\vert w_x^TAw_y - \frac{c}{N}N_xN_y\right\vert \leq \theta\sqrt{\sum_{i=1}^{N-1}\alpha_i^2\sum_{i=1}^{N-1}\beta_i^2} \leq \theta\sqrt{(w_x^T\widetilde{Z}\widetilde{Z}^Tw_x)
(w_y^T\widetilde{Z}\widetilde{Z}^Tw_y)}
$$
which by $\widetilde{Z}\widetilde{Z}^T=NI-uu^T$ and $w_x^T(NI-uu^T)w_x=N_x(N-N_x)$ proves \eqref{eq_lemma}. 
\hfill$\square$
\\
\\
A second proof for Lemma \ref{L_general isoperimetric inequality} can be formulated based Haemers' interlacing theorem and applications \cite{Haemers_1995}. Haemers ingeniously describes how quotient matrix constructions combined with the interlacing theorem can lead to algebraic expressions (i.e. involving Laplacian eigenvalues) for combinatorial quantities (i.e. possible number of links between subsets of nodes in a graph).
%-------
\\ \textbf{Proof B:}\\
Haemers defines the block-matrix $B$
\begin{equation}\label{construction Laplacian}
B = \begin{bmatrix}0 & Q+cI \\ Q+cI & 0 \end{bmatrix}
\end{equation}
for some graph Laplacian $Q$, and any scalar $c\in\mathbb{R}$. By the anti-diagonal blockform of $B$, we know that each eigenvalue $\mu_j$ of the Laplacian $Q$ corresponds to two eigenvalues $\tilde{\lambda_{i}} = \mu_j+c$ and $\tilde{\lambda}_{2N-i} = -(\mu_j+c)$ of $B$. 
\\
We consider a specific partitioning $\pi$ of the rows of $B$ (nodes in the combined graph), for which we can explicitly write the quotient matrix. For the Laplacian in the upper-right block, we partition the nodes $\mathcal{N}$ into a subset $\mathcal{N}_x$ of size $N_x$, and a remainder set $\mathcal{N}_{rx}$. For the Laplacian in the lower-left block, we partition the nodes $\mathcal{N}$ into a subset $\mathcal{N}_y$ of size $N_y$, where $\mathcal{N}_y$ is non-overlapping with the $N_x$-size block of the other Laplacian, and a remainder set $\mathcal{N}_{ry}$. Overall, this results in the partitioning $\lbrace{\mathcal{N},\mathcal{N}}\rbrace\rightarrow\left\lbrace\mathcal{N}_x,\mathcal{N}_{rx},\mathcal{N}_y,\mathcal{N}_{ry}\right\rbrace$ for matrix $B$. For this partitioning, and because $Bu=cu$ due to $Qu=0$, we can write the quotient matrix $B^{(\pi)}$ explicitly as:
\begin{equation} \label{quotient matrix Laplacian}
B^{(\pi)} = \begin{bmatrix}\frac{1}{N_x} &0&0&0 \\ 0&\frac{1}{N-N_x}&0&0 \\ 0&0&\frac{1}{N-N_y}&0 \\ 0&0&0&\frac{1}{N_y} \end{bmatrix}
\begin{bmatrix}0&0&cN_x+m&-m \\ 0&0&c(N-N_x-N_y)-m&cN_y+m \\ cN_x+m&c(N-N_x-N_y)-m&0&0 \\ -m&cN_y+m&0&0 \end{bmatrix},
\end{equation}
where $m$ is the number of links between subsets $\mathcal{N}_x$ and $\mathcal{N}_y$, i.e. $w_x^TAw_y$ in Lemma \ref{L_general isoperimetric inequality}. 
\\
We can write the determinant of $B^{(\pi)}$ in two ways: an equality involving $m$ and an inequality involving the eigenvalues of the Laplacian $Q$. Combining both expressions for the determinant then yields the isoperimetric inequality \eqref{eq_lemma} of Lemma \ref{L_general isoperimetric inequality}.
\\
From \eqref{quotient matrix Laplacian}, the determinant of $B^{(\pi)}$ can be calculated as:
\begin{equation} \label{quotient determinant 1}
\det\left(B^{(\pi)}\right) = \frac{c^2\left(cN_xN_y+Nm\right)^2}{N_x(N-N_x)N_y(N-N_y)}
\end{equation}
Secondly, if we call $\delta_1\geq\delta_2\geq\delta_3\geq\delta_4$ the eigenvalues of $B^{(\pi)}$, where $\delta_1=-\delta_4$ and $\delta_2=-\delta_3$ hold because of the anti-diagonal blockmatrix structure, then we have a second equation for the determinant:
\begin{equation} 
\det\left(B^{(\pi)}\right) = \delta_1\delta_2\delta_3\delta_4 = \delta_1^2\delta_2^2
\end{equation}
From the definition of $B^{(\pi)}$, it follows that the all-one vector $u$ is an eigenvector with eigenvalue $c$, i.e. $B^{(\pi)}u = cu$. This means that either $\vert\delta_1\vert=c$ or $\vert\delta_2\vert=c$. Additionally, because $B^{(\pi)}$ is a quotient matrix of $B$, we know that the eigenvalue sequence $\delta_i$ of $B^{(\pi)}$ interlaces the eigenvalue sequence $\tilde{\lambda}_i$ of $B$:
$$
-\tilde{\lambda}_2\leq\delta_1\leq\tilde{\lambda}_1 \text{ and } 
-\tilde{\lambda}_3\leq\delta_2\leq\tilde{\lambda}_2
$$
Because we know that either $\delta_1$ or $\delta_2$ equals $\tilde{\lambda}_i=\mu_N+c=c$, we can write:
\begin{equation} \label{quotient determinant 2}
\det\left(B^{(\pi)}\right) = \delta_1^2\delta_2^2 \leq c^2\left(\max_{\forall i\neq N}\vert\mu_i+c\vert\right)
\end{equation}
Combining \eqref{quotient determinant 1} and \eqref{quotient determinant 2} gives:
\begin{equation}
\frac{c^2\left(cN_xN_y+Nm\right)}{N_x(N-N_x)N_y(N-N_y)} \leq c^2\theta^2,
\end{equation}
with $\vert c+\mu_i\vert\leq\theta, \forall i\neq N$. By taking the square root of both sides, replacing $m$ by $w_x^TAw_y$ and $c$ by $-c$, we find again the isoperimetric inequality \eqref{eq_lemma} in Lemma \ref{L_general isoperimetric inequality}.
\hfill$\square$ \\ \\
\textit{Remark:} Proofs A and B are two different ways to arrive at the same result. Proof A, based on Chung's approach, involves two approximations that upper-bound the cut-set approximation. The first approximation is upper bounding the $(c-\mu_i)$ values by $\theta$, i.e. $\left\vert\sum_{i=1}^{N-1}\alpha_i\beta_i(c-\mu_i)\right\vert \leq \theta\sum_{i=1}^{N-1}\vert \alpha_i\beta_i\vert$. The second approximation involves the Cauchy-Schwartz inequality applied to the inner-product $\sum_{i=1}^{N-1}\vert \alpha_i\beta_i\vert$. Proof B based on Haemers' approach, involves one approximation step. The absolute value of the second largest eigenvalue $\vert\delta_2\vert$ of the quotient matrix $B^{(\pi)}$ is upper bounded by the second largest absolute eigenvalue $\max_{i\neq N}\vert \mu_i+c\vert$ of $Q+cI$, based on the interlacing theorem. 
\\
Since both approaches lead to the same result, we can conclude that the error due to interlacing is of the same nature as the error due to upper-bounding $(c-\mu_i)$ combined with the Cauchy-Schwartz inequality, which is a non-trivial relationship.
%---------------------------------------------------------------------
\subsection{Proof of Theorem \ref{topological approx theorem}}
Theorem \ref{topological approx theorem} follows from Lemma \ref{L_general isoperimetric inequality} by particular choices of $(c,A,w_x,w_y)$. 
\\
\textbf{Proof:}
\\
First, we choose $w_x=(u-w)\circ s_k$ and $w_y=w\circ s_m$, where $(.\circ s_k)$ represents the Hadamard product (elementwise product) with $s_k$. For this choice of $w_x$ and $w_y$, which are Bernoulli vectors satisfying $w_x^Tw_y=0$, and any adjacency matrix $A$, we can write:
$$
w_x^TAw_y = (u_k\circ(u-w))^TA(w\circ u_m) = (u-w)^TA^{(km)}w
$$
Secondly, we choose the specific value $c=N\tilde{a}_{km}$ which satisfies the condition $c\in\mathbb{R}$.
\\
These choices allow us to rewrite Lemma \ref{L_general isoperimetric inequality} as:
$$
\left\vert{(u-w)^TA^{(km)}w - (\tilde{u}-\tilde{w})^T\widetilde{A}^{(km)}\tilde{w}}\right\vert\leq
\frac{\theta}{N}\sqrt{\tilde{w}_m(N-\tilde{w}_m)(N_k-\tilde{w}_k)(N-(N_k-\tilde{w}_k))}
$$
for any adjacency matrix $A$, which equals equation \eqref{topological approx general graph} and thus proves Theorem \ref{topological approx theorem}.
\hfill$\square$ 
%---------------------------------------------------------------------
\subsection{Proof of Theorem \ref{bi-regular approx theorem}}
Theorem \ref{bi-regular approx theorem} states that the topological approximation error can be bounded more tightly for bi-regular graphs $A_{km,r}$, which we prove based on Haemers' interlacing techniques \cite{Haemers_1995}.
\\
\textbf{Proof:}
\\
Consider a bi-regular graph $G_{km,r}$ with partitions $\mathcal{N}_k$ and $\mathcal{N}_m$, for which the adjacency matrix has the block-form:
$$
A_{km,r} = \begin{bmatrix}0 & B \\ B^T & 0 \end{bmatrix},
$$
with $Bu = d_1u$ and $u^TB=d_2u^T$, because the graph is biregular. The values $d_1 = \frac{L}{N_k}$ and $d_2=\frac{L}{N_m}$ are the degrees of the partitions. \\
Furthermore, we consider a partitioning $\pi$ of the nodes of $G_{km,r}$ into four sets $\left\lbrace \mathcal{N}_k^x,\mathcal{N}_k^r,\mathcal{N}_m^y,\mathcal{N}_m^r \right\rbrace$ according to
$$
\begin{cases}
\mathcal{N}_k^x\cup\mathcal{N}_k^r = \mathcal{N}_k,~  \mathcal{N}_k^x\cap\mathcal{N}_k^r = \emptyset
\text{ and }
\vert\mathcal{N}_k^x\vert = N_x \\
\mathcal{N}_m^x\cup\mathcal{N}_m^r = \mathcal{N}_m,~ \mathcal{N}_m^x\cap\mathcal{N}_m^r = \emptyset 
\text{ and }
\vert\mathcal{N}_m^x\vert = N_x
\end{cases}
$$
In other words, partition $k$ is further refined into a subset of $N_x$ nodes and a remainder subset, and similarly for partition $m$. For this partitioning $\pi$, the quotient matrix can be explicitly written as:
\begin{equation}\label{eq_quotient bi-regular}
A_{km,r}^{(\pi)} = \begin{bmatrix}\frac{1}{N_x} &0&0&0 \\ 0&\frac{1}{N_k-N_x}&0&0 \\ 0&0&\frac{1}{N_y}&0 \\ 0&0&0&\frac{1}{N_m-N_y} \end{bmatrix}
\begin{bmatrix}0&0& m &\frac{L}{N_k}N_x-m \\ 0&0&\frac{L}{N_m}N_y-m&L(1-\frac{N_x}{N_k}-\frac{N_y}{N_m})+m \\ m &\frac{L}{N_m}N_y-m&0&0 \\ \frac{L}{N_k}N_x-m&L(1-\frac{N_x}{N_k}-\frac{N_y}{N_m})+m&0&0 \end{bmatrix}
\end{equation}
where $m$ is the number of links between partitions $\mathcal{N}_k^x$ and $\mathcal{N}_m^y$, i.e. the cut-set size $(u-w)^TA^{(km)}w$ in Theorem \ref{bi-regular approx theorem}.
\\
We can write the determinant of $A_{km,r}^{(\pi)}$ in two ways: an expression involving $m$, which follows directly from the block-matrix form and secondly, an inequality involving the eigenvalues of $A_{km,r}^{(\pi)}$. Combining both expressions for the determinant yields the isoperimetric inequality of Theorem \ref{bi-regular approx theorem}. From \eqref{eq_quotient bi-regular}, the determinant of $A_{km,r}^{(\pi)}$ can be calculated as:
\begin{equation} \label{determinant c}
\det\left(A_{km,r}^{(\pi)}\right) = \frac{L^2\left(m-\frac{L}{N_kN_m}N_xN_y\right)^2}{N_x(N_k-N_x)N_y(N_m-N_y)}
\end{equation}
Secondly, if we call $\delta_1\geq\delta_2\geq\delta_3\geq\delta_4$ the eigenvalues of $A_{km,r}^{(\pi)}$, where $\delta_1=-\delta_4$ and $\delta_2=-\delta_3$ hold because of the anti-diagonal block structure, then we have a second equation for the determinant:
\begin{equation} \label{determinant eigenvalues}
\det\left(A_{km,r}^{(\pi)}\right) = \delta_1\delta_2\delta_3\delta_4 = \delta_1^2\delta_2^2
\end{equation}
Next, two facts about the eigenvalues of $A_{km,r}^{(\pi)}$ are combined to find expression \eqref{topological approx bi-regular graph}. Firstly, because $A_{km,r}^{(\pi)}$ is a quotient matrix of $A_{km,r}$, we know by Theorem \ref{theorem_quotient interlacing} that the eigenvalues of the first interlace those of the latter. In other words, we can bound $\delta_2$ by:
\begin{align*}
\lambda_{N-K+2}\leq\delta_2\leq\lambda_2 
\end{align*}
Because $\lambda_{N-k+2}=\lambda_{N-2}=-\lambda_{3}\geq -\lambda_2$, we find 
\begin{equation} \label{first fact}
\delta_2^2\leq\lambda_2^2
\end{equation}
The second fact we use is 
\begin{equation}\label{second fact}
\delta_1=\frac{L}{\sqrt{N_kN_m}}
\end{equation}
which can be verified by considering the eigenvalue equation: 
$$
\left(A^{(\pi)}_{km,r}-\frac{L}{\sqrt{N_kN_m}}\right)\begin{bmatrix}\sqrt{N_m}\\\sqrt{N_m}\\\sqrt{N_k}\\\sqrt{N_k} \end{bmatrix} = 0
$$
from which follows that $\left[\sqrt{N_m},\sqrt{N_m},\sqrt{N_k},\sqrt{N_k}\right]^T$ is the right eigenvector of $A_{km,r}^{(\pi)}$ according to eigenvalue $\delta_1=\frac{L}{N_kN_m}$. By the Perron-Frobenius theorem \cite{Van_Mieghem_GS}, we know that for non-negative matrices such as $A_{km,r}^{(\pi)}$, the largest (possibly non-unique) eigenvalue accords to an eigenvector with non-negative elements. This means that $\delta_1$ is the largest eigenvalue of $A_{km,r}^{(\pi)}$ since its corresponding eigenvector is a vector with non-negative elements.
\\
Combining \eqref{first fact} and \eqref{second fact} then yields an upper-bound for the determinant of $A^{(\pi)}_{km,r}$ in equation \eqref{determinant eigenvalues}:
$$
\det\left(A_{km,r}^{(\pi)}\right) \leq \frac{L^2}{N_kN_m}\lambda_2^2
$$
Combined with \eqref{determinant c} this gives
$$
\frac{L^2\left(m-\frac{L}{N_kN_m}N_xN_y\right)^2}{N_x(N_k-N_x)N_y(N_m-N_y)} \leq \frac{L^2}{N_kN_m}\lambda_2^2
$$
Which reduces to equation \eqref{topological approx bi-regular graph} if we replace $m$ by $(u-w)^TA^{(km)}w$, and which thus proves Theorem \ref{bi-regular approx theorem}.
\hfill$\square$